\documentstyle[aps,]{revtex}
\draft
\begin{document}

\title{
Duality for symmetric second rank tensors. (I) : the massive
case.}
\author{H. Casini $^{a}$, R. Montemayor $^b$ and Luis F. Urrutia$^c$}
\address{
$^a$Centre die Physique Th\`eorique, Campus de Luminy,\\ F-13288
Marseille, France \\
$^b$Instituto Balseiro and CAB, Universidad Nacional de Cuyo and CNEA, \\
8400 Bariloche, Argentina\\
$^c$Departamento de F\'{\i}sica de Altas Energ\'{\i}as, Instituto
de Ciencias Nucleares \\Universidad Nacional Aut\'onoma de
M\'exico, A.P. 70-543, M\'exico D.F. 04510, M\'exico}
\maketitle

\begin{abstract}
A generalization of  duality transformations for arbitrary
Lorentz tensors is presented, and a systematic scheme for
constructing the dual descriptions is developed. The method, a
purely Lagrangian approach, is based on a first order parent
Lagrangian, from which the dual partners are generated. In
particular, a family of theories which are dual to the massive
spin two Fierz-Pauli field $h_{\mu\nu}$, both free and coupled to
an external source, is constructed in terms of a $T_{(\mu \nu
)\sigma }$ tensor.
\end{abstract}

\section{Introduction}

There usually is  a great deal of freedom in the choice of
variables for the description of a physical system. Different
choices of variables are considered equivalent when they are able
to describe the same system, and any difference is mainly due to
subjective choices. However, there might be practical reasons to
prefer a given description to others. For example, in some cases
it might be desirable to have a formulation where some symmetries
are made explicit in the Lagrangian. This usually requires the
use of a redundant set of variables to describe the system
configurations, as in the case of gauge theories. Conversely, in
other situations it is more convenient to choose a minimal,
non-redundant, set of variables. Another interesting example of
this freedom is the bosonization of fermionic systems. The actual
proof of the equivalence between different descriptions is
usually a non-trivial task.

 Duality, in its wider meaning, refers
to two equivalent descriptions for a physical system using
different fields.\cite{DUAL} One of the simplest cases is the
scalar-tensor duality. It corresponds to the equivalence between
a free massless scalar field $\varphi $, with field strength
$f_{\mu }=\partial _{\mu }\varphi $, and a massless antisymmetric
field $B_{\mu \nu }$, the Kalb-Ramond field, with field strength
$H_{\mu \nu \sigma }=\partial _{\mu }B_{\nu \sigma }+\partial
_{\nu }B_{\sigma \mu }+\partial _{\sigma }B_{\mu \nu }$
\cite{FLUND,KR,DSH}. Another example is in fact a predecessor of
the modern approach to duality, the electric-magnetic symmetry
$({\vec{E}}+i{\vec{B}})\rightarrow e^{i\phi }({
\vec{E}}+i{\vec{B}})$ of the free Maxwell equations. When there
are charged sources this symmetry can be maintained by
introducing magnetic monopoles \cite{MILTON}. This transformation
provides a connection between weak and strong couplings via the
Dirac quantization condition. At the level of Yang-Mills theories
with spontaneous symmetry breaking this kind of duality is
expected, due to the existence of topological dyon-type solitons
\cite{OLIVE}. The extension of electromagnetic duality to
$SL(2,Z)$ is usually referred to as S-duality, and plays an
important role in the non-perturbative study of field and string
theories \cite{GOMEZ}.

These basic ideas have been subsequently generalized to arbitrary
forms in arbitrary dimensions. Well known dualities are the ones
between massless $p$-form and $(d-p-2)$-form fields and between
massive $p$ and $(d-p-1)$-forms in $d$ dimensional
space-time\cite{FQUEV1}. These dualities among free fields have
been proved by using the method of parent Lagrangians \cite{PL} as
well as the canonical formalism \cite{LOZ}. They can be extended
to include source interactions \cite{QUEV2}.

The above duality among forms can be understood as a relation
between fields in different representations of the Lorentz group.
The origin of this equivalence can be traced using the little
group technique for constructing the representations of the
Poincar\'e group in $d$ dimensions. A detailed discussion of this
observation suggests the possibility of generalizing the duality
transformations among $p$-forms to tensorial fields with
arbitrary Young symmetry types. Consistent massless free
\cite{FRMIX}, interacting \cite{AG}, and massive \cite{t0}
theories of mixed Young symmetry tensors were constructed in the
past, but the attempts to prove a dual relation between these
descriptions did not lead to a positive answer \cite{t0}.
Additional interest in this type of theories arises from the
recent formulation of $d=11$ dimensional supergravity as a gauge
theory for the ${\rm osp}(32|1)$ superalgebra. It includes a
totally antisymmetric fifth-rank Lorentz tensor one form $b_{\mu
}{}^{abcde}$, whose mixed symmetry piece does not have any
related counterpart in the standard $\,d=11$ supergravity theory
\cite{JZ}.

In this paper we present a general scheme to construct dual
theories based on a Lagrangian approach. The method, sketched in
a preceding article \cite{plb}, has been originally motivated by
the relationship between field representations corresponding to
associated Young diagrams. Here we fully develop this approach on
a purely Lagrangian basis, and apply it to the case of a massive
spin-2 theory.

Let us consider a simple example, the scalar field $\varphi$, in
order to illustrate our general procedure for constructing dual
theories. The starting point is the corresponding second order
Lagrangian
\begin{equation}
L(\varphi )=\frac{1}{2}\partial _{\mu }\varphi \partial ^{\mu
}\varphi - \frac{1}{2}m^{2}\varphi ^{2}+J\varphi \text{\thinspace
},  \label{lff}
\end{equation}
where $m$ is the mass parameter of $\varphi $ and $J$ is a source
coupled to the field. As the first step, we construct a first
order Lagrangian, using a generalization of a procedure used in
Ref. \cite{Lanczos}. We are interested in a particular Lagrangian
structure, which we will call the standard form
\begin{equation}
L(\varphi ,L^{\mu })=L^{\mu }\partial _{\mu }\varphi -\frac{1}{2}L^{\mu
}L_{\mu }-\frac{1}{2}m^{2}\varphi ^{2}+J\varphi \,.  \label{llf}
\end{equation}
This standard form is defined by the kinetic term. It contains
the derivative of the original field times a new auxiliary
variable, which we call, in a rather loose way, the field strength
of the original theory. In this first order formulation  the
original field and the auxiliary field we have just introduced
define the configuration space. The equations of motion are
\begin{eqnarray}
m^{2}\varphi =-\partial _{\mu }L^{\mu }+J, \qquad L_{\mu } =\partial _{\mu
}\varphi \,.  \label{efl}
\end{eqnarray}

The key recipe to construct the dual theory is to introduce a point
transformation in the configuration space for the auxiliary variable
\begin{equation}
L^{\mu }=\epsilon ^{\mu \nu \sigma \tau }H_{\nu \sigma \tau }\,,
\label{DUALT}
\end{equation}
which leads to a new first order Lagrangian
\begin{equation}
L(\varphi ,H_{\nu \sigma \tau })=H_{\nu \sigma \tau }\epsilon ^{\mu \nu
\sigma \tau }\partial _{\mu }\varphi +3H_{\nu \sigma \tau }H^{\nu \sigma
\tau }-\frac{1}{2}m^{2}\varphi ^{2}+J\varphi \,.  \label{lhf}
\end{equation}
This turns out to be the parent Lagrangian from which both dual
theories can be obtained. In fact, using the equation of motion
for $H_{\nu \sigma \tau }$ we obtain $H_{\nu \sigma \tau
}(\varphi )$ which takes us back to our starting action
(\ref{lff}) after it is substituted in Eq. (\ref{lhf}). On the
other hand, we can also eliminate the field
 $\varphi $  from the Lagrangian using its equation of motion
\begin{equation}
m^{2}\varphi =-\partial _{\mu }\epsilon ^{\mu \nu \sigma \tau }H_{\nu \sigma
\tau }+J.  \label{efh}
\end{equation}
In such a way we now obtain the new theory
\begin{equation}
L(H_{\nu \sigma \tau })=\frac{1}{2}\left( \epsilon ^{\mu \nu \sigma \tau
}\partial _{\mu }H_{\nu \sigma \tau }\right) ^{2}+3m^{2}H_{\nu \sigma \tau
}H^{\nu \sigma \tau }-J\epsilon ^{\mu \nu \sigma \tau }\partial _{\mu
}H_{\nu \sigma \tau }+\frac{1}{2}J^{2}\,,  \label{lhh}
\end{equation}
which is equivalent to the original one
through the transformation (\ref{efh}).
In this form we have obtained a theory  dual to (\ref{lff}).

For a massless theory, $m=0$, we lose the connection between the
original field $\varphi$ and the new one $H_{\nu \sigma \tau}$.
In this case Eq. (\ref{efh}) becomes a constraint on $H_{\nu
\sigma \tau }$
\begin{equation}
\partial _{\mu }\epsilon ^{\mu \nu \sigma \tau }H_{\nu \sigma \tau }=J\,.
\end{equation}
Out of the sources, the above equation tells us that the field
$H_{\nu \sigma \tau}$ can be considered as a field strength with
an associated potential.

Another paradigmatic example of dualization is the standard
S-duality for electrodynamics with a $\theta$ term. In the
following we describe a method to deal with this case which,
together with the previous example, will serve as a motivation for
the general scheme to be  presented in the next section.  Let us
consider the Euclidean Lagrangian
\begin{equation}
L=\frac{1}{8\pi }\left( \frac{4\pi }{g^{2}}F_{\mu \nu }F^{\mu \nu
}+\frac{ i\,\theta }{2\pi }\frac{1}{2}\epsilon _{\mu \nu \rho
\sigma }F^{\mu \nu }F^{\rho \sigma }\right).  \label{EED}
\end{equation}
Using the notation
\begin{equation}
\tau =\frac{\theta }{2\pi }+\frac{4\pi i
}{g^{2}},\;\;\bar{\tau}=\frac{ \theta }{2\pi }-\frac{4\pi i
}{g^{2}}\;, \label{DUALN}
\end{equation}
the standard Euclidean S-dualization recipe
\begin{equation}
F\rightarrow {\tilde F},\quad {\tilde {\tilde F}} \rightarrow +F,
\quad \bar{\tau}\rightarrow \frac{1}{\bar{\tau}},\;\;\tau
\rightarrow \frac{1}{ \tau }
\end{equation}
leads to a new Lagrangian
\begin{equation}
\tilde{L}=-\frac{1}{8\pi }\left( \frac{4\pi }{g^{2}\bar{\tau}\tau }\tilde{F}
_{\mu \nu }\tilde{F}^{\mu \nu }-\frac{i}{\bar{\tau}\tau }\frac{\theta }{2\pi
}\frac{1}{2}\epsilon _{\mu \nu \rho \sigma }\tilde{F}^{\mu \nu }\tilde{F}
^{\rho \sigma }\right).  \label{DLED}
\end{equation}
For the purpose of our discussion it is more convenient to write
the initial Lagrangian (\ref {EED}) as
\begin{equation}
L=\frac{1}{8\pi }\left( aF_{\mu \nu }F^{\mu \nu }+ib\frac{1}{2}\epsilon
_{\mu \nu \rho \sigma }F^{\mu \nu }F^{\rho \sigma }\right)  \label{EED1},
\end{equation}
where
\begin{equation}
a=\frac{4\pi }{g^{2}},\;\;b=\frac{\theta }{2\pi },\;\;\bar{\tau}\tau
=a^{2}+b^{2}.
\end{equation}
In this notation, the Euclidean dual is
\begin{equation}
\tilde{L}=-\frac{1}{8\pi }\left(
\frac{a}{a^{2}+b^{2}}\tilde{F}_{\mu \nu } \tilde{F}^{\mu \nu
}-\frac{ib}{a^{2}+b^{2}}\frac{1}{2}\epsilon _{\mu \nu \rho \sigma
}\tilde{F}^{\mu \nu }\tilde{F}^{\rho \sigma }\right).
\label{DLED1}
\end{equation}
Now we will show how to go from Lagrangian (\ref{EED}) to
Lagrangian (\ref{DLED}) using the basic ideas of our approach. To
begin with, we construct a first order Lagrangian for (\ref
{EED1}), introducing the Lagrange multiplier $ G^{\mu\nu}$
\begin{eqnarray}
L(F,A,G) =\frac{1}{8\pi}\left(
aF_{\mu\nu}F^{\mu\nu}+ib\;\frac{1}{2}
\epsilon_{\mu\nu\rho\sigma}F^{\mu\nu}F^{\rho\sigma}\right)
 -\frac{1}{4\pi}\left( G^{\mu\nu}F_{\mu\nu}-G^{\mu\nu}\left( \partial
_{\mu}A_{\nu}-\partial_{\nu}A_{\mu}\right) \right).  \label{EED2}
\end{eqnarray}
The Euler-Lagrange equation for $F_{\mu\nu}$
\begin{equation}
aF_{\mu\nu}+ib\frac{1}{2}\epsilon_{\mu\nu\rho\sigma}F^{\rho\sigma}=G_{\mu\nu}
\end{equation}
leads to
\begin{equation}
F^{\alpha\beta}=\frac{1}{\left( a^{2}+b^{2}\right) }\left(
aG^{\alpha\beta}-ib\frac{1}{2}G_{\mu\nu}
\epsilon^{\mu\nu\alpha\beta}\right) \label{DEFEG}
\end{equation}
by a purely algebraic manipulation. This allows us to eliminate
this field from Lagrangian (\ref{EED2}), obtaining
\begin{eqnarray}
 L(A,G)=-\frac{1}{8\pi}\frac{1}{\left( a^{2}+b^{2}\right) }\left(
aG^{\alpha\beta}-ib\frac{1}{2}\epsilon^{\alpha\beta\mu\nu}G_{\mu\nu}\right)
G_{\alpha\beta} +\frac{1}{4\pi}G^{\mu\nu}\left(
\partial_{\mu}A_{\nu}-\partial_{\nu}A_{\mu}\right)
\label{FOLAG3}
\end{eqnarray}
which identifies $G^{\mu\nu}$ as the field strength of $A^\mu$.
The above first order Lagrangian is equivalent to the second order
Lagrangian (\ref{EED1}). This can be verified via the solution
\begin{equation}
G^{\alpha\beta}=a\left(
\partial^{\alpha}A^{\beta}-\partial^{\beta}A^{\alpha }\right)
+ib\frac{1}{2}\left( \partial_{\rho}A_{\sigma}-\partial_{\sigma
}A_{\rho}\right) \epsilon^{\rho\sigma\alpha\beta}  \label{POTREL1}
\end{equation}
of the equation of motion for $G_{\mu\nu}$,
together with the definition (\ref{DEFEG}). The variation of
 $A_\mu$ in Lagrangian (\ref{FOLAG3}) produces the remaining equation
\begin{equation}
\partial_\mu G^{\mu\nu}=0.
\label{EQG2}
\end{equation}
Now we define  the dual field $H_{\alpha\beta}$
\begin{equation}
G^{\mu\nu}=\frac{1}{2}\epsilon^{\mu\nu\rho\sigma}H_{\rho\sigma}.
\end{equation}
By substitution  in  Eq. (\ref{FOLAG3}) we obtain
\begin{equation}
L=-\frac{1}{8\pi}\frac{1}{\left( a^{2}+b^{2}\right) }\left(
aH_{\kappa
\lambda}-ib\frac{1}{2}\epsilon_{\kappa\lambda\rho\sigma}H^{\rho\sigma}
\right) H^{\kappa\lambda}
+\frac{1}{4\pi}\frac{1}{2}\epsilon_{\mu\nu\rho\sigma}H^{\mu\nu}(\partial^\rho
A^\sigma-\partial^\sigma A^\rho), \label{DUALTE}
\end{equation}
which is the correspondent parent Lagrangian. The variation of
this last Lagrangian with respect to $A^\mu$ produces a Bianchi
identity for $H_{\rho\sigma}$
\begin{equation}
\epsilon^{\nu\mu\rho\sigma}\partial_{\mu}H_{\rho\sigma}=0,
\end{equation}
which implies
\begin{equation}
H_{\rho\sigma}(B)=\partial_{\rho}B_{\sigma}-\partial_{\sigma}B_{\rho},
\end{equation}
where $H_{\rho\sigma}$ is identified as the dual field strength.
Using this property in Eq. (\ref{DUALTE}) leads to the second
order Lagrangian
\begin{equation}
L(B)=-\frac{1}{8\pi}\frac{1}{\left( a^{2}+b^{2}\right) }\left(
aH_{\kappa
\lambda}-ib\frac{1}{2}\epsilon_{\kappa\lambda\rho\sigma}H^{\rho\sigma}
\right) H^{\kappa\lambda},
\end{equation}
the dual version of the original one. The above Lagrangian is
precisely (\ref{DLED1}) with the  notation $H^{\mu\nu}={\tilde
F}^{\mu\nu}$. The relation with the original theory appears at
the level of the potentials  and is given by
\begin{equation}
a\left(
\partial^{\alpha}A^{\beta}-\partial^{\beta}A^{\alpha}\right) +
\frac{ib}{2}\left(
\partial_{\rho}A_{\sigma}-\partial_{\sigma}A_{\rho }\right)
\epsilon^{\rho\sigma\alpha\beta}=\frac{1}{2}\epsilon^{\alpha
\beta\rho\sigma}\left(
\partial_{\rho}B_{\sigma}-\partial_{\sigma}B_{\rho }\right).
\end{equation}
The case $b=0$ reduces to standard Electrodynamics and leads to
\begin{eqnarray}
&&L(A)=\frac{a}{8\pi}F_{\mu\nu}F^{\mu\nu},\;\;\;F_{\rho\sigma}=\left(
\partial_{\rho}A_{\sigma}-\partial_{\sigma}A_{\rho}\right), \\
&&L(B)=-\frac{1}{8\pi a }H_{\kappa\lambda}H^{\kappa\lambda
},\;\;H_{\rho\sigma}=\left(
\partial_{\rho}B_{\sigma}-\partial_{\sigma
}B_{\rho}\right)=a{\tilde F}_{\rho\sigma},
\end{eqnarray}
together with the relation
\begin{equation}
a\left( \partial^{\alpha}A^{\beta}-\partial^{\beta}A^{\alpha}\right) =\frac{1
}{2}\epsilon^{\alpha\beta\rho\sigma}\left( \partial_{\rho}B_{\sigma
}-\partial_{\sigma}B_{\rho}\right).
\end{equation}

This paper focuses on the construction of a  dual theory for a
massive spin-2 field in four dimensions. It is organized as
follows. In Section II we formulate the general scheme for
dualization pursued here. Section III contains the construction
of an  auxiliary first order Lagrangian which is equivalent to the
usual one in terms of the standard Fierz-Pauli field $h_{\mu\nu}$
for a massive spin-2 particle. The general method for constructing
such an auxiliary Lagrangian is briefly reviewed in Appendix A.
An explicit proof of the equivalence between this auxiliary
Lagrangian and the massive Fierz-Pauli Lagrangian is given in
Appendix B. Section IV contains the definition of the dual field
$T_{(\mu\nu)\sigma}$ together with the  construction of the
parent Lagrangian. In Section V the duality transformations
arising from the parent Lagrangian are derived. The dual
Lagrangian, in terms of $T_{(\mu\nu)\sigma}$, is obtained in
Section VI together with the corresponding equations of motion
and the set of Lagrangian contraints. These contraints allow us
to make sure that we have obtained the correct number of degrees
of freedom. Most of the calculations in this section are
relegated to Appendix C. In Section VII we discuss the example of
a fixed point mass $m$ whose field is calculated in each of  the
dual theories, thus allowing the explicit verification of the
duality trasformations. Finally we close with Section VIII which
contains a summary of the work together with some comments
regarding preliminary work in the zero mass limit of the present
approach. A complete discussion of the massless case is deferred
to a forthcoming publication.

\section{The dualization procedure}

In general terms, the method applied to the previous examples can
be summarized as follows, assuming that there is no external
source, for simplicity. We start from a second order theory for
the free field\ $\Phi $ of a given tensorial character, which can
be schematically presented as
\begin{equation}
L(\Phi )=\frac{1}{2}\partial \Phi \partial \Phi
-\frac{M^{2}}{2}\Phi \Phi \,. \label{LGSO}
\end{equation}
Next, we introduce an auxiliary field $W$ to construct a first
order formulation in the standard form
\begin{equation}
L(\Phi ,W)=\left( \partial \Phi \right)
W-\frac{1}{2}WW-\frac{M^{2}}{2}\Phi \Phi \,, \label{LGFO}
\end{equation}
as explained in Appendix A. This identifies $W$ as the field
strength of $\Phi$, with the equation of motion
\begin{equation}
\partial W+M^{2}\Phi =0\,.  \label{FSEM}
\end{equation}

Now, we introduce $\tilde{W}$ as the field strength dual to $W$
via the change of variables
\begin{equation}
W=\epsilon \;\tilde{W}\,,
\end{equation}
and substitute in the first order action (\ref{LGFO}) to obtain the Lagrangian
\begin{equation}
{\tilde L}(\Phi ,\tilde{W})=\left(
\partial \Phi \right) \epsilon \tilde{W}-\frac{1}{2}\epsilon
\tilde{W}\epsilon \tilde{W}-\frac{M^{2}}{2}\Phi \Phi \,.
\label{PLGEN}
\end{equation}

In this way we obtain the parent Lagrangian (\ref{PLGEN}) which
generates the pair of dual theories. In fact, the field
$\tilde{W}$ can always be eliminated from Lagrangian (\ref{PLGEN})
to recover the initial Lagrangian (\ref {LGSO}).

The Euler-Lagrange equation for $\Phi$ is
\begin{equation}
\epsilon \partial \tilde{W}+M^{2}\Phi =0\,.  \label{FCE}
\end{equation}
If $M\neq 0$, or more generally if it is a regular matrix, Eq.
(\ref{FCE}) allows the algebraic elimination of the field $\Phi $
in Lagrangian (\ref{PLGEN}), yielding a second order Lagrangian
for $\tilde{W}$
\begin{equation}
{\tilde L}\left(\Phi \equiv -\frac{1}{M^{2}}\epsilon \partial
\tilde{W}\,,\,\tilde{W}\right) \propto \frac{1}{2}\epsilon
\partial \tilde{W}\epsilon \partial
\tilde{W}-\frac{M^2}{2}\epsilon\tilde{W}\epsilon\tilde{W},
\end{equation}
which is the dual to the original $L(\Phi )$.

If $M=0$, the parent Lagrangian reduces to
\begin{equation}
{\tilde L}(\Phi ,\tilde{W})=\left(
\partial \Phi \right) \epsilon \tilde{W}-\frac{1}{2}\epsilon
\tilde{W}\epsilon \tilde{W}\,,  \label{PLM01}
\end{equation}
with the equations of motion
\begin{eqnarray}
&& \epsilon \partial \tilde{W}=0\,,  \label{EM1M0} \\
&& \epsilon \tilde{W}\epsilon -\epsilon \left( \partial \Phi \right) =0\,,
\label{EM2M0}
\end{eqnarray}
preventing  the algebraic solution for $\Phi$. Nevertheless, Eq.
(\ref{EM1M0}) is a Bianchi identity for $\tilde{W}$ whose solution
can be written symbolically as
\begin{equation}
\tilde{W}=\partial B\,.  \label{DGF}
\end{equation}
That is to say, the dual field $\tilde{W}$ is a field strength
and can be written in terms of a new potential $B$. The dual
Lagrangian results from substituting Eq. (\ref{DGF}) into
Lagrangian (\ref{PLM01}) and is
\begin{equation}
L(B)=\frac{1}{2}\epsilon \partial B\epsilon \partial B\,,
\end{equation}
where we have explicitly used the Bianchi identity in the second
term of the RHS of Eq. (\ref{PLM01}). Finally, the relation
\begin{equation}
\epsilon \partial B\epsilon -\epsilon \partial \Phi =0,\,
\end{equation}
obtained from Eq. (\ref{EM2M0}), provides the connection between
the dual theories.

\section{Massive Fierz-Pauli Lagrangian}

The Lagrangian for the massive Fierz-Pauli field is
\begin{eqnarray}
\mathcal{L} &=&-\partial _{\mu }h^{\mu \nu }\partial _{\alpha }h_{\nu
}^{\alpha }+\frac{1}{2}\partial _{\alpha }h^{\mu \nu }\partial ^{\alpha
}h_{\mu \nu }+\partial _{\mu }h^{\mu \nu }\partial _{\nu }h_{\alpha
}^{\alpha }-\frac{1}{2}\partial _{\alpha }h_{\mu }^{\mu }\partial ^{\alpha
}h_{\nu }^{\nu }  \nonumber \\
&&-\frac{M^{2}}{2}\left( h_{\mu \nu }h^{\mu \nu }-h_{\mu }^{\mu }h_{\nu
}^{\nu }\right) +h_{\mu \nu }\Theta ^{\mu \nu },  \label{FLP}
\end{eqnarray}
where $\Theta ^{\mu \nu }$ is the source described by a symmetric
tensor, not necessarily conserved in contrast to the massless
case. The kinetic part of Lagrangian (\ref{FLP}) is just the
linearized Einstein Lagrangian. The equations of motion are
\begin{eqnarray}
\partial ^{\alpha }\partial _{\alpha }h_{\mu \nu }+\partial _{\mu }\partial
_{\nu }h_{\alpha }^{\alpha }-g_{\mu \nu }\left( \partial ^{\beta }\partial
_{\beta }h_{\alpha }^{\alpha }-\partial _{\alpha }\partial _{\beta
}h^{\alpha \beta }\right)-\left( \partial _{\mu }\partial _{\alpha
}h_{\nu }^{\alpha }+\partial _{\nu }\partial _{\alpha }h_{\mu }^{\alpha
}\right)
+M^{2}\left( h_{\mu \nu }-g_{\mu \nu }h_{\alpha }^{\alpha }\right) =\Theta
_{\mu \nu }.  \label{em}
\end{eqnarray}
Taking the trace and the divergence of this equation we have
\begin{eqnarray}
&&h_{\alpha }^{\alpha } =-\frac{1}{3M^{2}}\left( \Theta _{\alpha }^{\alpha }-
\frac{2}{M^{2}}\partial _{\alpha }\partial _{\beta }\Theta ^{\alpha \beta
}\right) \,,  \label{a1} \\
&&\partial^{\mu }h_{\mu \nu } =\frac{1}{M^{2}}\left( \partial _{\mu }\Theta
_{\nu }^{\mu }-\frac{1}{3}\partial _{\nu }\Theta _{\alpha }^{\alpha }+\frac{2
}{3M^{2}}\partial _{\nu }\partial _{\alpha }\partial _{\beta }\Theta
^{\alpha \beta }\right) \,,  \label{a2}
\end{eqnarray}
which show that the trace and the divergence of $h_{\mu\nu}$ do
not propagate, vanishing outside the sources, as expected for a
pure spin-2 theory. Eq. (\ref{em}) now becomes
\begin{equation}
\left( \partial ^{\alpha }\partial _{\alpha }+M^{2}\right) h_{\mu \nu }=
\tilde{\Theta}_{\mu \nu }\,\,,
\end{equation}
where $\tilde{\Theta}_{\mu \nu }$ is a source dependent term that
can be expressed in terms of $\Theta _{\mu \nu }$ and its
derivatives.

Following the procedure sketched in Appendix A, we can construct
an equivalent first order Lagrangian in the standard form. This
Lagrangian is not unique, because of the freedom in the choice of
the auxiliary fields. Alternatively, we can construct a
Lagrangian having the standard form with arbitrary coefficients,
which are subsequently adjusted to obtain the original Lagrangian
when the auxiliary fields are eliminated. In the present case the
last approach is simpler, and we will follow it. Therefore, we
start by proposing a  field strength $K^{\alpha \left\{ \beta
\sigma \right\}}$ satisfying the following symmetry properties
\begin{eqnarray}
&&K^{\alpha \left\{ \beta \sigma \right\} }=K^{\alpha \left\{ \sigma \beta
\right\} }\,,\label{FSSP0} \\
&&K^{\alpha \left\{ \beta \sigma \right\} }+K^{\beta \left\{ \sigma \alpha
\right\} }+K^{\sigma \left\{ \alpha \beta \right\} }=0\,.  \label{FSSP}
\end{eqnarray}
These symmetry properties greatly simplify the manipulations and,
as it will become evident in the following, they are consistent
with the degrees of freedom of the spin-2 massive field. With
this auxiliary field we construct the  first order Lagrangian
\begin{eqnarray}
L &=&-\frac{1}{6}aK^{\alpha \{\beta \sigma \}}K_{\alpha \{\beta
\sigma \}}+ \frac{1}{8}qK^{\beta }K_{\beta }-\frac{2}{9}r\epsilon
_{\;\;\;}^{\gamma \delta \kappa \lambda }K_{\kappa \{\lambda
}{}^{\sigma \}}K_{\gamma
\{\delta \sigma \}}  \nonumber \\
&&-\frac{e}{\sqrt{2}}K^{\alpha \{\beta \sigma \}}\partial _{\alpha
}h_{\sigma \beta }-\frac{M^{2}}{2}\left( h_{\mu \nu }h^{\mu \nu }-h_{\mu
}{}^{\mu }h_{\nu }{}^{\nu }\right) +\Theta _{\mu \nu }h^{\mu \nu }  \nonumber \\
&&+\Lambda _{\alpha \{\beta \sigma\} }\left( K^{\alpha \{\beta \sigma
\}}+K^{\beta \{\sigma \alpha \}}+K^{\sigma \{\alpha \beta \}}\right) \,,
\label{FOLFP}
\end{eqnarray}
where $K_{\alpha }=K_{\alpha \{\lambda }{}^{\lambda \}}$. This
Lagrangian has the most general mass term for the field
$K^{\alpha \left\{ \beta \sigma \right\} }$ with the symmetry
properties (\ref{FSSP0}-\ref{FSSP}). Here $K^{\alpha \left\{
\beta \sigma \right\} }$ is identified as the field strength of
$h_{\sigma \beta }$. The constraint (\ref{FSSP}) is enforced by
the Lagrange multiplier$\;\Lambda _{\alpha \{\beta \sigma \}} =
\Lambda _{\alpha \{ \sigma \beta\}}$.  In Appendix B we show that
the elimination of $K^{\alpha \left\{ \beta \sigma \right\} }$
and $\Lambda _{\alpha \{ \sigma \beta\}}$ in (\ref{FOLFP}) leads
effectively to the Fierz-Pauli Lagrangian  when the coefficients
satisfy
\begin{eqnarray}
4r^{2} &=&a\left( e^{2}-a\right) \,\,,  \label{cr1} \\
3q &=&2a+e^{2}\,.  \label{cr2}
\end{eqnarray}
In such a case both theories are equivalent and Lagrangian
(\ref{FOLFP}) is the first order standard Lagrangian for the
Fierz-Pauli massive field. From conditions (\ref{cr1}-\ref{cr2})
only two independent coefficients in Lagrangian (\ref{FOLFP})
remain, one of them being the normalization of the auxiliary
field.

\section{Dual field and parent Lagrangian}

Now that we have identified the field strength $K^{\alpha \left\{
\beta \sigma \right\} }$ for $h_{\mu\nu}$ and the corresponding
first order theory, we can implement the transformation
\begin{equation}
K^{\alpha \left\{ \beta \sigma \right\} }\rightarrow \Omega
_{\left( \mu \nu \xi \right) }^{\left\{ \beta \sigma \right\}
},\;\;\;\;\;\;K^{\alpha \left\{ \beta \sigma \right\} }=\epsilon
^{\alpha \mu \nu \xi }\Omega _{\left( \mu \nu \xi \right)
}^{\left\{ \beta \sigma \right\} }  \label{DTDEF}
\end{equation}
that leads to the dual theory. Substituting  this transformation
in Eq. (\ref{FOLFP}), we obtain the parent Lagrangian
\begin{eqnarray}
L&=& a\Omega _{\left( \mu \nu \xi \right) }^{\left\{ \beta \sigma \right\}
}\Omega _{\left\{ \beta \sigma \right\} }^{\left( \mu \nu \xi \right) }-
\frac{1}{4}\left( 2a+e^{2}\right) \Omega _{\left( \mu \nu \xi
\right) }\Omega ^{\left( \mu \nu \xi \right)
}+\frac{2}{3}\sqrt{a\left( e^{2}-a\right) }\epsilon _{\;\rho \tau
\chi }^{\mu }\Omega _{\left( \mu \nu \lambda \right) }^{\left\{
\lambda \sigma \right\} }\Omega _{\;\;\;\;\;\;\;\;\;\sigma
\}}^{\left( \rho \tau \chi \right) \{\nu }  \nonumber
\\
&& -\frac{e}{\sqrt{2}}\epsilon ^{\alpha \mu \nu \xi }\Omega _{\left( \mu \nu
\xi \right) }^{\left\{ \beta \sigma \right\} }\partial _{\alpha }h_{\sigma
\beta }-\frac{M^{2}}{2}\left( h_{\mu \nu }h^{\mu \nu }-h_{\mu }{}^{\mu
}h_{\nu }{}^{\nu }\right) +\Theta _{\mu \nu }h^{\mu \nu }  \nonumber \\
&& +\Lambda _{\alpha \beta \sigma }\left( \epsilon ^{\alpha \mu \nu \xi
}\Omega _{\left( \mu \nu \xi \right) }^{\left\{ \beta \sigma \right\}
}+\epsilon ^{\beta \mu \nu \xi }\Omega _{\left( \mu \nu \xi \right)
}^{\left\{ \sigma \alpha \right\} }+\epsilon ^{\sigma \mu \nu \xi }\Omega
_{\left( \mu \nu \xi \right) }^{\left\{ \alpha \beta \right\} }\right),
\label{DLAGFO}
\end{eqnarray}
where
\begin{equation}
\Omega _{\left( \mu \nu \xi \right) }=g_{\alpha \beta }\Omega _{\left( \mu
\nu \xi \right) }^{\left\{ \alpha \beta \right\} }\,.
\end{equation}
The dual theory is derived by eliminating $h_{\sigma \beta }$.
Alternatively, by eliminating $\Omega _{\left\{ \beta \sigma
\right\} }^{\left( \mu \nu \xi \right) }$ from Eq. (\ref{DLAGFO})
we recover the Fierz-Pauli theory. The field $\Omega _{\left\{
\beta \sigma \right\} }^{\left( \mu \nu \xi \right) }$  satisfies
the constraint
\begin{equation}
\epsilon ^{\alpha \mu \nu \xi }\Omega _{\left( \mu \nu \xi \right)
}^{\left\{ \beta \sigma \right\} }+\epsilon ^{\beta \mu \nu \xi }\Omega
_{\left( \mu \nu \xi \right) }^{\left\{ \sigma \alpha \right\} }+\epsilon
^{\sigma \mu \nu \xi }\Omega _{\left( \mu \nu \xi \right) }^{\left\{ \alpha
\beta \right\} }=0\,,  \label{CO}
\end{equation}
as a consequence of Eq. (\ref{FSSP}). A simple way to warrant
this constraint is to express  $\Omega _{\left( \rho \sigma \tau
\right) }^{\left\{ \beta \gamma \right\} }$ in terms of a tensor
$T_{\left( \rho \sigma \right) }^{\gamma }=-T_{\left( \sigma \rho
\right) }^{\gamma }$, as follows:
\begin{equation}
\Omega _{\left( \rho \sigma \tau \right) }^{\left\{ \beta \gamma \right\} }=-
\frac{1}{3\sqrt{2}}\left( g_{\tau }^{\beta }T_{\left( \rho \sigma \right)
}{}^{\gamma }+g_{\rho }^{\beta }T_{\left( \sigma \tau \right) }{}^{\gamma
}+g_{\sigma }^{\beta }T_{\left( \tau \rho \right) }{}^{\gamma }+g_{\tau
}^{\gamma }T_{\left( \rho \sigma \right) }{}^{\beta }+g_{\rho }^{\gamma
}T_{\left( \sigma \tau \right) }{}^{\beta }+g_{\sigma }^{\gamma }T_{\left(
\tau \rho \right) }{}^{\beta }\right) \,.  \label{ot}
\end{equation}
This expression  identically satisfies the constraint, and avoids
the necessity of its explicit use throughout the remaining
manipulations. The duality transformation (\ref{DTDEF}) now reads
\begin{equation}
K^{\alpha\{ \beta\sigma\} }=-\frac{1}{\sqrt{2}}\left( T_{\left(
\mu\nu\right) }{}^{\sigma}\epsilon^{\mu\nu\alpha\beta}+T_{\left(
\mu \nu\right) }{}^{\beta}\epsilon^{\mu\nu\alpha\sigma}\right)
\label{KET}
\end{equation}
with
\begin{equation}
K^{\alpha}=-\frac{1}{2}K^{\alpha\{\beta}{}_{\beta\}}=-\sqrt{2}
\epsilon^{\mu\nu\alpha\beta}T_{\left( \mu\nu\right) \beta}.
\end{equation}
The trace of $T_{\left( \mu \nu \right) \beta }$ does not
contribute to the expression (\ref{ot}). Thus, we will take
$T_{\left( \mu \nu \right) \beta }$ to be traceless and impose
this constraint by means of a Lagrange multiplier. The  analysis
in Section VI will show that this choice is indeed compatible
with the dynamics of the Fierz-Pauli field.

Using the identities
\begin{eqnarray}
&&\epsilon _{\;\;\alpha \beta }^{\mu \nu }T_{(\mu \nu )\sigma }\left(
T^{(\alpha \beta )\sigma }+2T^{(\sigma \alpha )\beta }\right) =-\frac{2}{3}
T_{(\mu \nu )}^{\nu }\epsilon ^{\alpha \beta \gamma \mu }\left(
T_{(\alpha \beta )\gamma }+T_{(\gamma \alpha )\beta }+T_{(\beta
\gamma )\alpha },\right) \\
&&\epsilon ^{\mu \nu \alpha \beta }T_{(\mu \nu )\sigma
}T_{(\alpha \beta )}{}^{\sigma }=-2\epsilon _{\alpha }^{\;\;\mu
\nu \beta }T_{(\mu \nu )\sigma }^{\;\;\;\;}T^{(\sigma \alpha
)}{}_\beta=-4\epsilon _{\alpha }^{\;\mu \nu \beta }T_{(\mu \sigma
)\nu }T^{(\sigma \alpha )}{}_\beta, \label{ID}
\end{eqnarray}
which follows from the antisymmetry of $T^{(\mu \nu )\sigma }$
and the null trace property $T^{(\mu \nu )}{}_{\nu }=0$, we
rewrite the parent Lagrangian (\ref{DLAGFO}) as
\begin{eqnarray}
&&L=\frac{1}{3}\left( 2a-\frac{1}{2}e^{2}\right) T_{\left( \mu
\nu \right) }{}^{\sigma }T^{\left( \mu \nu \right)
}{}_\sigma+\frac{1}{3}\left( 2a+e^{2}\right) T_{\left( \mu \nu
\right) \beta }T^{\left( \mu \beta \right) \nu
}+\frac{1}{2}\sqrt{a\left( e^{2}-a\right) }\epsilon ^{\mu \nu
\kappa \lambda }T_{\left( \mu \nu \right) }{}^{\sigma }T_{\left(
\kappa \lambda
\right) \sigma }  \nonumber \\
&&+eT_{\left( \mu \nu \right) }{}^{\sigma }\epsilon ^{\mu \nu
\alpha \beta }\partial _{\alpha }h_{\sigma \beta
}-\frac{M^{2}}{2}\left( h_{\mu \nu }h^{\mu \nu }-h_{\mu }{}^{\mu
}h_{\nu }{}^{\nu }\right) +\Theta _{\mu \nu }h^{\mu \nu }+\lambda
_{\beta }T^{(\beta\alpha)}{}_\alpha\,.  \label{parent}
\end{eqnarray}
The above Lagrangian is equivalent to Eq. (12) in Ref. \cite{plb}
when $\Theta_{\mu\nu}=0$.

\section{Duality transformations}

Varying  $h^{\mu \nu }$ in the parent Lagrangian
(\ref{parent}) we obtain the Euler-Lagrange equation
\begin{equation}
M^{2}\left( h_{\alpha }{}^{\alpha }g^{\mu \nu }-h^{\mu \nu
}\right) +\Theta ^{\mu \nu }-\frac{e}{2}\left( \partial _{\sigma
}T_{(\alpha \beta )}^{\mu }\epsilon ^{\alpha \beta \sigma \nu
}+\partial _{\sigma }T_{(\alpha \beta )}^{\nu }\epsilon ^{\alpha
\beta \sigma \mu }\right) =0\,.
\end{equation}
From here we  compute the trace
\begin{equation}
\;\;h_{\alpha }{}^{\alpha }=\frac{1}{3M^{2}}\left( e\epsilon
^{\alpha \beta \sigma \nu }\partial _{\sigma }T_{(\alpha \beta
)\nu }-\Theta _{\alpha }^{\alpha }\right) \,,
\end{equation}
which allows to solve for $h_{\mu\nu}$
\begin{equation}
h^{\mu \nu }=-\frac{e}{2M^{2}}\left( \epsilon ^{\alpha \beta
\sigma \nu }\partial _{\sigma }T_{(\alpha \beta )}{}^{\mu
}+\epsilon ^{\alpha \beta \sigma \mu }\partial _{\sigma
}T_{(\alpha \beta )}{}^{\nu }\right) +\frac{1}{ 3M^{2}}g^{\mu \nu
}\left( e\epsilon ^{\alpha \beta \rho \kappa }\partial _{\rho
}T_{(\alpha \beta )\kappa }-\Theta _{\alpha }^{\alpha }\right)
+\frac{ 1}{M^{2}}\Theta ^{\mu \nu }\,,  \label{EQMH}
\end{equation}
giving the first  duality relation $h^{\mu\nu}=h^{\mu\nu}(T)$.

Varying  $T^{(\mu \nu )\sigma }$ we derive the equation
\begin{eqnarray}
\frac{1}{3}\left( 4a-e^{2}\right) T^{(\mu \nu )\sigma } &+&\frac{1
}{3}\left( 2a+e^{2}\right) \left( T^{(\mu \sigma )\nu }-T^{(\nu \sigma )\mu
}\right)
+\sqrt{a\left( e^{2}-a\right) }\;\epsilon ^{\mu \nu \alpha \beta
}T_{(\alpha \beta )}{}^{\sigma } \nonumber \\
&+&e\epsilon ^{\mu \nu \alpha \beta }\partial _{\alpha
}h_{\;\beta }^{\sigma }+\frac{1}{2}\left( \lambda ^{\mu }g^{\nu
\sigma }-\lambda ^{\nu }g^{\mu \sigma }\right) =0\,.  \label{EQMT}
\end{eqnarray}
From here, contracting with the metric and the Levi-Civita tensors  we have
\begin{eqnarray}
&& T^{(\mu \nu)}{}_\nu=-\frac{1}{2a-e^{2}}\frac{e}{a}\left[
\sqrt{a\left( e^{2}-a\right) }\left( \partial _{\sigma }h^{\sigma
\mu }-\partial ^{\mu }h\right) +\frac{3
}{4}e\lambda ^{\mu }\right],  \label{5} \\
&& T^{(\mu \nu )\sigma }+T^{(\nu \sigma )\mu }+T^{(\sigma \mu )\nu }=\frac{1}{
2a-e^{2}}\epsilon ^{\mu \nu \sigma \lambda }\left[ e\left( \partial _{\sigma
}h_{\;\lambda }^{\sigma }-\partial _{\lambda }h\right) +\frac{3}{2a}\sqrt{
a\left( e^{2}-a\right) }\lambda _{\lambda }\right],\\
&&a\epsilon _{\mu \nu \kappa \lambda }T^{(\mu \nu )\sigma }
=2\sqrt{a\left( e^{2}-a\right) }\;T_{(\kappa \lambda )}{}^{\sigma
}+e\left( \partial _{\kappa }h_{\;\lambda }^{\sigma }-\partial
_{\lambda }h_{\;\kappa }^{\sigma }\right) +\frac{1}{2}\epsilon
_{\;\kappa \lambda \mu }^{\sigma }\lambda ^{\mu }
\nonumber \\
&&\qquad \qquad \qquad
-\frac{1}{3}\frac{2a+e^{2}}{2a-e^{2}}\left( g_{\kappa }^{\sigma
}g_{\lambda }^{\rho }-g_{\lambda }^{\sigma }g_{\kappa }^{\rho
}\right) \left[ e\left( \partial _{\sigma }h_{\;\rho }^{\sigma
}-\partial _{\rho }h\right) + \frac{3}{2a}\sqrt{a\left(
e^{2}-a\right) }\lambda _{\rho }\right] , \label{9}
\end{eqnarray}
after some algebraic manipulations. Using these relationships we
solve for $T^{(\mu \nu )\sigma }$ in Eq. (\ref{EQMT}), obtaining
\begin{eqnarray}
T^{(\mu \nu )\sigma }&=&-\frac{1}{2e}\epsilon ^{\mu \nu \alpha \beta
}\partial _{\alpha }h_{\;\beta }^{\sigma }+\frac{1}{6e}\frac{2a+e^{2}}{
2a-e^{2}}\epsilon ^{\mu \nu \sigma \lambda }\left( \partial _{\sigma
}h_{\;\lambda }^{\sigma }-\partial _{\lambda }h\right)
 -\frac{1}{2ae}\sqrt{a\left( e^{2}-a\right) }\left( \partial ^{\mu
}h^{\sigma \nu }-\partial ^{\nu }h^{\sigma \mu }\right)  \nonumber \\
&& +\frac{1}{6ae}\frac{2a+e^{2}}{2a-e^{2}}\sqrt{a\left( e^{2}-a\right) }\;
\left[ g^{\sigma \mu }\left( \partial _{\kappa }h^{\kappa \nu }-\partial
^{\nu }h\right) -g^{\sigma \nu }\left( \partial _{\kappa }h^{\kappa \mu
}-\partial ^{\mu }h\right) \right]  \nonumber \\
&& +\frac{1}{2a}\frac{1}{2a-e^{2}}\sqrt{a\left( e^{2}-a\right)
}\epsilon ^{\mu \nu \sigma \lambda }\lambda _{\lambda
}-\frac{e^{2}}{4a}\frac{1}{ 2a-e^{2}}\left( \lambda ^{\mu }g^{\nu
\sigma }-\lambda ^{\nu }g^{\mu \sigma }\right),  \label{EQMTT}
\end{eqnarray}
which constitutes the second duality relation
$T^{(\mu\nu)\sigma}=T^{(\mu\nu)\sigma} (h, \lambda)$. Only the
Lagrange multiplier $\lambda ^{\mu } $ remains to be varied,
which imposes the null trace constraint on $T^{(\mu \nu )\sigma
}$. Summarizing, using only algebraic manipulations and without
any mixing between the results of different variations, the
Lagrangian equations of motion are (\ref {EQMH}) and (\ref{EQMTT})
together with
\begin{equation}
T^{(\mu \nu )}{}_{\nu }=0.  \label{lm}
\end{equation}
Eqs. (\ref {EQMH}) and  (\ref{EQMTT}) are those to be used in
eliminating either $ h^{\sigma \mu }$ or $T^{(\mu \nu )\sigma }$
from the parent Lagrangian, to obtain the corresponding
Lagrangians for $T^{(\mu \nu )\sigma }$ or $h^{\sigma \mu }$
respectively.

In fact, the degrees of freedom of $h^{\sigma \mu }$ are mapped into the
traceless part of $T^{(\mu \nu )\sigma }$, $\hat{T}^{(\mu \nu )\sigma }$.
The relationship between the Fierz-Pauli field and $\hat{T}^{(\mu \nu
)\sigma }$ can be obtained in a straightforward way as follows. The null trace
condition imposes
\begin{equation}
\lambda ^{\mu }=- \frac{4}{3e} \sqrt{a\left( e^{2}-a\right) }\left( \partial
_{\sigma }h^{\sigma \mu }-\partial ^{\mu }h\right).
\end{equation}
Using this constraint to eliminate the Lagrange multiplier in Eq. (\ref{EQMTT})
we get
\begin{eqnarray}
 \hat{T}^{(\mu \nu )\sigma }&=&-\frac{1}{2e}\epsilon ^{\mu \nu \alpha \beta
}\partial _{\alpha }h_{\;\beta }^{\sigma }+\frac{1}{2e}\epsilon ^{\mu \nu
\sigma \lambda }\left( \partial _{\sigma }h_{\;\lambda }^{\sigma }-\partial
_{\lambda }h\right) -\frac{1}{2ae}\sqrt{a\left( e^{2}-a\right) }\left(
\partial ^{\mu }h^{\sigma \nu }-\partial ^{\nu }h^{\sigma \mu }\right)
\nonumber \\
&& +\frac{1}{6ae}\sqrt{a\left( e^{2}-a\right) }\;\left[ g^{\sigma \mu }\left(
\partial _{\kappa }h^{\kappa \nu }-\partial ^{\nu }h\right) -g^{\sigma \nu
}\left( \partial _{\kappa }h^{\kappa \mu }-\partial ^{\mu
}h\right) \right],
\end{eqnarray}
which is the final expression for the duality transformation
$\hat{T}^{(\mu \nu )\sigma }=\hat{T}^{(\mu \nu )\sigma }(h)$.

All previous relations greatly simplify on shell. Under this
circumstance the condition
\begin{equation}
\left( \partial ^{\kappa }h_{\kappa }^{\lambda }-\partial
^{\lambda }h_{\kappa }^{\kappa }\right) =\frac{1}{M^{2}}\partial
_{\eta }\Theta ^{\eta \lambda }\,
\end{equation}
for the trace and divergence of $h^{\sigma \lambda }$ is
obtained, using Eqs. (\ref{a1}) and (\ref{a2}). The remaining on
shell  constraints are
\begin{eqnarray}
&&T_{(\mu \nu )}{}^{\mu }=0\,, \\
&&\partial _{\beta }T_{(\mu \,\nu )}{}^{\beta }=-\frac{\sqrt{a\left(
e^{2}-a\right) }}{3aeM^{2}}\left( \partial _{\mu }\partial _{\alpha }\Theta
_{\nu }^{\alpha }-\partial _{\nu }\partial _{\alpha }\Theta _{\mu }^{\alpha
}\right) \,, \\
&&\epsilon ^{\lambda \mu \nu \beta }T_{(\mu \nu )\beta }=\frac{2}{eM^{2}}
\partial _{\mu }\Theta ^{\mu \lambda }\,, \\
&&\partial _{\kappa }T^{(\kappa \lambda )\sigma }
+\frac{1}{2a}\sqrt{a\left( e^{2}-a\right) }\epsilon _{\mu \nu
}{}^{\kappa \lambda }\partial _{\kappa }T^{(\mu \nu )\sigma }=
-\frac{(2a+e^{2})}{6aeM^{2}}\epsilon _{\nu \mu
}^{\;\;\,\,\,\,\lambda \sigma }\partial ^{\mu }\partial _{\beta
}\Theta ^{\nu \beta }  \nonumber \\
&&\qquad\qquad\qquad\qquad\qquad\qquad+\frac{\sqrt{ a\left(
e^{2}-a\right) }}{3eaM^{2}}\left( -\partial ^{\sigma }\partial
_{\mu }\Theta ^{\mu \lambda }+g^{\lambda \sigma }\partial _{\beta
}\partial _{\eta }\Theta ^{\eta \beta }\right) \,.  \label{CONS}
\end{eqnarray}

Using the constraint equations for both fields the on shell
duality relations become
\begin{eqnarray}
&&h^{\mu \nu }=-\frac{e}{2M^{2}}\left( \epsilon ^{\alpha \beta \sigma \nu
}\partial _{\sigma }T_{(\alpha \beta )}{}^{\mu }+\epsilon ^{\alpha \beta
\sigma \mu }\partial _{\sigma }T_{(\alpha \beta )}{}^{\nu }\right) +\frac{1}{
3M^{2}}g^{\mu \nu }\left( \frac{2}{M^{2}}\partial _{\gamma }\partial
_{\lambda }\Theta ^{\gamma \lambda }-\Theta _{\alpha }^{\alpha }\right) +
\frac{1}{M^{2}}\Theta ^{\mu \nu }\,, \\
&&T_{(\mu \nu )\beta }=-\frac{1}{2e}\epsilon _{\,\,\,\,\,\,\,\mu \nu
}^{\alpha \sigma }\partial _{\alpha }h_{\sigma \beta }-\frac{\sqrt{a\left(
e^{2}-a\right) }}{2ae}\left( \partial _{\mu }h_{\nu \beta }-\partial _{\nu
}h_{\mu \beta }\right)  \nonumber \\
&&\;\;\;\;\;\;\;\;\;\;\;\;\;\;-\frac{1}{2eM^{2}}\epsilon _{\,\,\,\,\mu \nu \beta
}^{\sigma }\partial _{\eta }\Theta _{\sigma }^{\eta }-\frac{\sqrt{a\left(
e^{2}-a\right) }}{6aeM^{2}}\left( g_{\nu \beta }\,\partial _{\eta }\Theta
_{\mu }^{\eta }-g_{\mu \beta }\,\partial _{\eta }\Theta _{\nu }^{\eta
}\right) \,.
\end{eqnarray}

In the particular case where $a=e^{2}$ the duality
transformations acquire the usual form
\begin{equation}
T_{(\mu \nu )\beta }=-\frac{1}{2e}\epsilon _{\,\,\,\,\,\,\,\mu \nu }^{\alpha
\sigma }\partial _{\alpha }h_{\sigma \beta }-\frac{1}{2eM^{2}}\epsilon
_{\,\,\,\,\mu \nu \beta }^{\sigma }\partial _{\eta }\Theta _{\sigma }^{\eta
}\,,
\end{equation}
involving only the Levi-Civita tensor. The constraint
(\ref{CONS})\ reduces to $\partial _{\kappa }T^{(\kappa \lambda
)\sigma }=0$ out of sources, which means that the field
$T^{(\kappa \lambda )\sigma }$ contains purely transversal
degrees of freedom. Otherwise, if $a\neq e^{2}$, the degrees of
freedom of $h^{\sigma \beta }$ are also mapped in the
longitudinal components $\left( \partial _{\kappa }T^{(\kappa
\sigma )\beta }+\partial _{\kappa }T^{(\kappa \beta )\sigma
}\right) $.

\section{Dual theory}

The substitution of $h_{\sigma \beta }(T)$, given by Eq.
(\ref{EQMH}), in the parent Lagrangian (\ref{parent}) leads to
the following Lagrangian for $T_{(\mu \nu )\sigma }$
\begin{eqnarray}
&&L=\frac{1}{3}\left( 2a-\frac{1}{2}e^{2}\right) T_{\left( \mu \nu \right)
\sigma}T^{\left( \mu \nu \right)\sigma }+\frac{1}{3}\left( 2a+e^{2}\right)
T_{\left( \mu \nu \right) \beta }T^{\left( \mu \beta \right) \nu }+\frac{1}{2
}\sqrt{a\left( e^{2}-a\right) }\epsilon ^{\mu \nu \kappa \lambda }T_{\left(
\mu \nu \right) }{}^{\sigma }T_{\left( \kappa \lambda \right) \sigma }
\nonumber \\
&&+\frac{1}{2}h_{\sigma \beta }(T)\left( -e\,\epsilon ^{\mu \nu
\alpha \beta }\partial _{\alpha }T_{\left( \mu \nu \right)
}{}^{\sigma }+\Theta ^{\sigma \beta }\right) +\lambda _{\beta
}T^{\beta }\,.
\end{eqnarray}
After some algebra and dropping a global $-e^{2}/(2M^{2})$ factor we obtain
\begin{eqnarray}
L &=&\frac{4}{9}F_{(\alpha \beta \gamma )\nu }\,F^{(\alpha \beta
\gamma )\nu }+\frac{2}{3}F_{(\alpha \beta \gamma )\nu
}\,F^{(\alpha \beta \nu )\gamma }-F_{(\alpha \beta \mu )}{}^{\mu
}\,F^{(\alpha \beta \nu )}{}_{\nu }  \nonumber
\\
&&-\frac{2M^{2}}{3e^{2}}\left[ \left( 2a-\frac{1}{2}e^{2}\right)
T_{(\mu \nu )\sigma }T^{(\mu \nu )\sigma }+\left( 2a+e^{2}\right)
T_{(\mu \nu )\sigma }T^{(\mu \sigma )\nu
}+\frac{3}{2}\sqrt{a\left( e^{2}-a\right) }\epsilon ^{\mu \nu
\alpha \beta }T_{(\mu \nu )\sigma }T_{(\alpha \beta )}{}^{\sigma }
\right]  \nonumber \\
&&+\lambda _{\beta }T^{\beta }+T_{(\alpha \beta )\mu}\,J^{(\alpha \beta
)\mu}.  \label{LAG}
\end{eqnarray}
Here we have introduced the field strength
\begin{equation}
F_{(\alpha \beta \gamma )\nu }=\partial _{\alpha }T_{(\beta
\gamma )\nu }+\partial _{\beta }T_{(\gamma \alpha )\nu }+\partial
_{\gamma }T_{(\alpha \beta )\nu },
\end{equation}
and the source term is given as a function of the source $\Theta
_{\mu \nu }$ by
\begin{equation}
J^{(\alpha \beta )\mu}=\frac{2}{e}\left(
\frac{1}{3}\epsilon^{\alpha \beta \rho \mu}\partial _{\rho
}\Theta _{\alpha }^{\alpha }-\epsilon ^{\alpha \beta \sigma \nu
}\partial _{\sigma }\Theta _{ \nu }{}^\mu\right) \,.  \label{J}
\end{equation}
Note that the new source $J^{(\alpha \beta )\mu}$ is traceless
\begin{equation}
J^{\alpha }=J^{(\alpha \beta )}{}_\beta=\frac{2}{e}\left(
\frac{1}{3} \epsilon _{\,\,\,\,\,\,\,\,\,\,\beta }^{\alpha \beta
\rho }\partial _{\rho }\Theta _{\alpha }^{\alpha }-\epsilon
^{\alpha \beta \sigma \nu }\partial _{\sigma }\Theta _{\beta \nu
}\right) =0,
\end{equation}
and also satisfies
\begin{eqnarray}
\epsilon _{\kappa \alpha \beta \mu }J^{(\alpha \beta )\mu } =-\frac{4}{e}
\partial _{\mu }\Theta^{\mu }{}_\kappa, \quad \partial _{\alpha }J^{(\alpha
\beta )\mu} =0, \quad \partial _{\mu }J^{(\alpha \beta )\mu} =-\frac{2}{e}
\epsilon ^{\alpha \beta \sigma \nu }\partial _{\sigma }\partial ^{\mu
}\Theta _{\mu \nu }.
\end{eqnarray}

As stated previously $T_{(\mu \nu )\sigma }=-$ $T_{(\nu \mu
)\sigma}$ and therefore the field $T_{(\nu \mu )\sigma }$ has 24
components. But not all of them are true degrees of freedom,
because there are cyclic variables. This becomes clear by defining
\begin{equation}
T_{(\mu \nu )\sigma }=\hat{T}_{(\mu \nu )\sigma }-\frac{1}{3}\left( g_{\mu
\sigma }T_{\nu }-g_{\nu \sigma }T_{\mu }\right)
\end{equation}
where $T_{\mu }\equiv $ $T_{(\mu \beta )}{}^{\beta }$, and
$\hat{T}_{(\mu \nu )\sigma }$ is a traceless field,
$\hat{T}_{(\mu \nu )}{}^{\nu }=0$. Next, we  rewrite Lagrangian
(\ref{LAG}) in terms of $\hat{T}^{(\chi \psi )\sigma }$ and
$T_{\mu }$ and we further use the Euler-Lagrangian equation for
$T_{\mu }$ to  eliminate this variable from the Lagrangian. The
resulting Lagrangian contains linear and bilinear terms in
$\lambda _{\beta }$. Finally, using the corresponding
Euler-Lagrangian equation for $\lambda _{\beta }$  we can also
eliminate this variable. In such a way we obtain an alternative
version of the dual Lagrangian for $\hat{T} _{(\mu \nu )\sigma }$
\begin{eqnarray}
L&=&\frac{4}{9}\hat{F}_{(\alpha \beta \gamma )\nu
}\,\hat{F}^{(\alpha \beta \gamma )\nu
}+\frac{2}{3}\hat{F}_{(\alpha \beta \gamma )\nu }\,\hat{F}
^{(\alpha \beta \nu )\gamma }-\hat{F}_{(\alpha \beta \mu
)}{}^{\mu }\,\hat{F} ^{(\alpha \beta \nu )}{}_{\nu
}+\hat{T}_{(\alpha \beta )}{}^{\mu }\,J_{\mu
}^{(\alpha \beta )}  \nonumber \\
&& -\frac{2M^{2}}{3e^{2}}\left[ \left( 2a-\frac{1}{2}e^{2}\right) \hat{T}
_{(\mu \nu )\sigma }\hat{T}^{(\mu \nu )\sigma }+\left( 2a+e^{2}\right) \hat{T
}_{(\mu \nu )\sigma }\hat{T}^{(\mu \sigma )\nu }\right.  \nonumber \\
&& \left. +\frac{1}{2}\left( e^{2}-a\right) \hat{T}_{(\mu \nu
)\sigma }\left( \hat{T}^{(\mu \nu )\sigma }+\hat{T}^{(\sigma \mu
)\nu }+\hat{T}^{(\nu \sigma )\mu }\right) +\frac{3}{2}
\sqrt{a\left( e^{2}-a\right) }\epsilon ^{\mu \nu \alpha \beta
}\hat{T}_{(\mu \nu )\sigma }\hat{T}_{(\alpha \beta )}{}^{\sigma }
\right].  \label{LAG2}
\end{eqnarray}
This clearly shows that the degrees of freedom are in the
traceless field $\hat{T}^{(\mu \nu )\sigma }$, as has already been
assumed in Section IV. When varying Lagrangian (\ref{LAG2}) it is
necessary to take into account that not all the components of
$\hat{T}^{(\mu \nu )\sigma }$ are independent, because of the
traceless condition, and this is rather cumbersome.

Consequently, in order to study the properties of the dual field
${T}^{(\mu \nu )\sigma }$ it is more convenient to go back to
Lagrangian (\ref{LAG}), because there we have to impose only the
antisymmetry constraint. In this way, starting from this
Lagrangian we obtain the equations of motion
\begin{eqnarray}
E^{(\beta\gamma)\nu}:&=& \frac{2}{3}\partial _{\alpha }\left[ 2\,F^{(\alpha
\beta \gamma )\nu }+\,\left( F^{(\alpha \beta \nu )\gamma }+F^{(\gamma
\alpha \nu )\beta }+F^{(\beta \gamma \nu )\alpha }\right) \right] -\partial
^{\nu }\,F^{(\beta \gamma \kappa )}{}_{\kappa }  \nonumber \\
&& -\partial _{\alpha }\left( g^{\gamma \nu }\,F^{(\alpha \beta \kappa
)}{}_{\kappa }+g^{\beta \nu }\,F^{(\gamma \alpha \kappa )}{}_{\kappa
}\right) +\frac{M^{2}}{e^{2}}\sqrt{a\left( e^{2}-a\right) }\epsilon ^{\beta
\gamma \kappa \lambda }T_{(\kappa \lambda )}{}^{\nu }  \nonumber \\
&& +\frac{2M^{2}}{3e^{2}}\left[ \left( 2a-\frac{1}{2}e^{2}\right) T^{(\beta
\gamma )\nu }+\frac{1}{2}\left( 2a+e^{2}\right) \left( T^{(\beta \nu )\gamma
}-T^{(\gamma \nu )\beta }\right) \right]  \nonumber \\
&& -\frac{1}{4}\left( \lambda ^{\beta }g^{\gamma \nu }-\lambda ^{\gamma
}g^{\beta \nu }\right) -\frac{1}{2}\,J^{(\beta \gamma )\nu }=0 , \label{EQM}\\
T^{(\mu\nu)}{}_\nu&=&0. \label{EQM1}
\end{eqnarray}

In order to make explicit the Lagrangian constraints arising from
Eq. (\ref {EQM}), which determine the number of propagating
degrees of freedom associated to the massive field
$T^{(\mu\nu)\rho}$, we consider the case of zero sources. In
Appendix C we provide some details of the derivation and the
results are summarized here. The constraints are
\begin{equation}
T^{(\mu \alpha )}{}_\alpha=0,\qquad
T^{(\mu \alpha) \beta}+T^{( \alpha\beta) \mu} + T^{(\beta \mu)\alpha}=0,
\;\;\qquad \partial _{\theta }T^{(\alpha \beta )\theta }=0,  \label{C1}
\end{equation}
\begin{equation}
0=\left( \partial _{\beta }T^{(\beta \gamma )\nu } + \partial _{\beta
}T^{(\beta \nu )\gamma } \right)+\sqrt{\frac{\left( e^{2}-a\right) }{a}}
\frac{1}{2}\left( \epsilon ^{\beta \gamma \kappa \lambda }\partial _{\beta
}T_{(\kappa \lambda )}{}^{\nu }+\epsilon ^{\beta \nu \kappa \lambda
}\partial _{\beta }T_{(\kappa \lambda )}{}^{\gamma
}\right):=S^{\{\gamma\nu\}},  \label{C2}
\end{equation}
where we observe that the antisymmetric part of $\partial _{\beta
}T^{(\beta \gamma )\nu }$ turns out to be zero, as shown in the
sourceless version of Eq. (\ref{SIMDT33}) of Appendix C. The count
of the number of independent degrees of freedom goes as follows.
The field $T^{(\mu\nu)\rho}$ has 24 independent components. Eqs.
(\ref{C1}) provide $4+4+6=14$ constraints respectively, thus
leaving 24-14=10 independent variables up to this level. Because
of the symmetry $S^{\gamma\nu}=S^{\nu\gamma}$, the remaining Eq.
(\ref{C2}) provides only $10$ relations. Nevertheless, among them
we find 5 additional identities: $4$ arising from
$\partial_\nu\,S^{\{ \gamma\nu\}} =0$ and $1$ arising from
$g_{\gamma\nu}S^{\{\gamma\nu\}}=0$, leaving only 5 additional
independent constraints. Thus, Eq. (\ref{C2}) reduces to $10-5=5$
the previous 10 independent degrees of freedom, as appropriate
for a massive spin-$2$ system.

Taking into account the constraints (\ref{C1}-\ref{C2}), the
equation of motion with zero sources becomes
\begin{eqnarray}
\partial ^{2}T^{(\beta \gamma )\nu }+\partial _{\alpha }\partial ^{\gamma
}T^{(\alpha \beta )\nu }-\partial _{\alpha }\partial ^{\beta
}T^{(\alpha \gamma )\nu }+\frac{M^{2}}{2e^{2}}\sqrt{a\left(
e^{2}-a\right) }\epsilon ^{\beta \gamma \kappa \lambda
}T_{(\kappa \lambda )}{}^{\nu } +\frac{M^{2}a}{ e^{2}}T^{(\beta
\gamma )\nu }=0. \label{mej0}
\end{eqnarray}
Furthermore, from Eq. (\ref{C2}) we can write
\begin{equation}
\partial _{\alpha }\partial _{\gamma }T^{(\alpha \beta )\nu }
-\partial _{\alpha }\partial _{\beta }T^{(\alpha \gamma )\nu }=
-\frac{D}{4a}\left[
\partial _{\gamma }\left( \epsilon ^{\alpha \beta \kappa \lambda
}\partial _{\alpha }T_{(\kappa \lambda )}{}^{\nu }+\epsilon
^{\alpha \nu \kappa \lambda }\partial _{\alpha }T_{(\kappa
\lambda )}{}^{\beta }\right)\ -\partial _{\beta }\left( \epsilon
^{\alpha \gamma \kappa \lambda }\partial _{\alpha }T_{(\kappa
\lambda )}{}^{\nu }+\epsilon ^{\alpha \nu \kappa \lambda }\partial
_{\alpha }T_{(\kappa \lambda )}{}^{\gamma }\right)\right] .
\end{equation}
The above equation and its dual imply
\begin{equation}
\left( \partial ^{\gamma }\partial _{\alpha }T^{(\alpha \nu )\beta
}-\partial ^{\beta }\partial _{\alpha }T^{(\alpha \nu )\gamma
}\right) =- \frac{D^{2}}{ae^{2}}\partial^2 T^{(\beta \gamma )\nu
}+\frac{D}{2e^{2}}\partial^2 \epsilon^{\beta
\gamma}{}_{\kappa\lambda} T^{(\kappa\lambda)\nu }.
\end{equation}
Hence, the equation of motion (\ref{mej0}) and its dual can be
written as
\begin{eqnarray}
\frac{a}{e^{2}}\partial^2 T+\frac{D}{e^{2}}
\partial^2 T^{*}+\frac{aM^{2}}{e^{2}}T+\frac{
DM^{2}}{e^{2}}T^{*} &=&0 \ ,\\
\frac{a}{e^{2}}\partial^2 T^{*}-\frac{D}{e^{2}}
\partial^2 T+\frac{aM^{2}}{e^{2}}T^{*}-\frac{ DM^{2}}{e^{2}}T
&=&0,
\end{eqnarray}
where we have omitted the indices of $T^{(\beta \gamma )\nu }$ and
$T^*$ means the dual of $T$. Finally,
 solving for $T^{(\beta \gamma )\nu }$ we
get
\begin{equation}
(\partial_2+M^{2})T^{(\beta \gamma )\nu }=0 \ .
\end{equation}

The simpler case $e^2=a$ is reminiscent of the standard duality
transformations  and the constraint equations simplify to
\begin{equation}
T^{(\mu \theta )}{}_\theta=0,\qquad T_{\;\;\;\;\;}^{(\mu \alpha
\beta )}=0, \qquad
\partial _{\theta }T^{(\mu\nu )\theta }=0, \qquad \partial
_{\theta}T^{(\theta \mu )\nu }=0 .
\end{equation}

\section{Point mass source}

As an illustration let us discuss a simple example: the field
generated by a point mass $m$ at rest.  The equation of motion in
the massive Fierz-Pauli theory, arising from Eqs. (\ref{em}),
(\ref{a1}) and (\ref{a2}), is
\begin{equation}
\left(  \partial^{\alpha}\partial_{\alpha}+M^{2}\right)  h_{\mu\nu}
=\Theta_{\mu\nu}-\frac{1}{3}\left(  g_{\mu\nu}+\frac{1}{M^{2}}\partial_{\mu
}\partial_{\nu}\right)  \Theta_{\alpha}^{\alpha}.
\end{equation}
The corresponding source is the energy-momentum tensor of a point
mass at rest and  has the components
\begin{eqnarray}
\Theta_{00}=16\pi m\delta({\mathbf{r}}),\qquad \Theta_{0i} =\Theta_{ji}=0.
\end{eqnarray}
The resulting field configuration is
\begin{eqnarray}
h_{00}=\frac{8}{3}\frac{m}{r}e^{-Mr},\qquad
h_{ij}=\frac{1}{2}\delta_{ij}h_{00}-\frac{1}{2M^{2}}\partial_{j}
\partial_{i}h_{00}, \qquad h_{0i}=0. \label{hmunu}
\end{eqnarray}
The last term in $h_{ij}$ is irrelevant when the field is coupled
to a conserved source.

It is interesting to observe  how the zero mass limit
discontinuity (the van Dam-Veltman-Zakharov discontinuity
\cite{VVS}-\cite{BD}) manifests itself here. In the  limit
$M\rightarrow0$, $h_{00}$ converges to $\frac{4}{3}$ of the
Newtonian potential $\frac{m}{r}$. Besides, the component $h_{ij}$
has a divergent term plus $\frac{1}{2}h_{00}\delta_{ij}$. This
has to be contrasted  with the massless spin-$2$ theory in the
Lorentz gauge, where the non-zero fields  are
\begin{eqnarray}
\tilde{h}_{00}=\frac{2m}{r}\qquad
\tilde{h}_{ij}=\tilde{h}_{00}\delta_{ij}+\partial_i\partial_j\,f(r),
\end{eqnarray}
where the last term in $\tilde{h}_{ij}$ accounts for a remaining
gauge freedom associated to a time independent spatial rotation.

Next we consider  the corresponding theory for
$T_{(\alpha\beta)\gamma}$, in the simpler case of  $a=e^{2}$. We
refer the reader to Appendix C for the notation.  Here the dual
source is
\begin{eqnarray}
&&J^{(\alpha\beta)\mu}=\frac{32\pi m}{e}\left(  g^{\mu0}\epsilon
^{0\alpha\beta\rho}-\frac{1}{3}\epsilon^{\alpha\beta\mu\rho}\right)
\partial_{\rho}\delta(\mathbf{r}),\label{ds}\\
&&J^{(\alpha\beta\mu)}=\frac{32\pi m}{e}\left(  g^{\mu0}\epsilon
^{0\alpha\beta\rho}+g^{\alpha0}\epsilon^{0\beta\mu\rho}+g^{\beta0}
\epsilon^{0\mu\alpha\rho}-\epsilon^{\alpha\beta\mu\rho}\right)  \partial
_{\rho}\delta(\mathbf{r}).
\end{eqnarray}
Our conventions are $\epsilon^{0123}=\epsilon_{123}=+1$. The
equation of motion, arising from Eq. (\ref{EQTE2A}) of Appendix
C,  becomes
\begin{equation}
\left(  \partial_{\alpha}\partial^{\alpha}+M^{2}\right)
T^{(\beta\gamma)\nu
}=\frac{1}{4}\,J^{(\beta\gamma)\nu}+\frac{1}{4}J^{(\gamma\beta\nu)}+\frac
{1}{6M^{2}}\partial_{\alpha}\partial^{\alpha}J^{\left(
\gamma\beta\nu\right) }, \label{eme}
\end{equation}
with the constraints
\begin{eqnarray}
 M^{2}T^{(\beta\gamma\nu)}=-\frac{1}{2}\,J^{(\beta\gamma\nu)},\qquad
 M^{2}\partial_{\theta}T^{(\mu\nu)\theta}=0,\qquad
M^{2}{\partial_{\beta}}T^{(\beta\gamma)\nu}=0.
\end{eqnarray}
From here, the non zero components of $T_{(\alpha\beta)\gamma}$ are
\begin{eqnarray}
&&T_{(0i)j}=\frac{2m}{3e}(1+Mr)\frac{e^{-Mr}}{r^{3}}\epsilon_{ijk}
x_{k}\label{h1},\\
&&T_{(ij)0}=-\frac{4m}{3e}(1+Mr)\frac{e^{-Mr}}{r^{3}}\epsilon_{ijk}
x_{k}.\label{h2}
\end{eqnarray}

We can now compare both theories. In terms of the massive
Fierz-Pauli solution, the solution for $T_{\left(  \mu\nu\right)
\sigma}$ can be written as
\begin{eqnarray}
&&T_{(0i)j}=-\frac{1}{4e}\epsilon_{ijk}\partial_{k}h_{00}\label{k1}, \\
&&T_{(ij)0}=+\frac{1}{2e}\epsilon_{ijk}\partial_{k}h_{00}\label{k2}.
\end{eqnarray}
It is straightforward to verify that both solutions, $h_{\mu\nu}$ and
$T_{(\alpha\beta)\gamma}$, are in fact related by the duality transformations. Out
of the source the on-shell duality relations are
\begin{eqnarray}
&&h_{\alpha\beta}=-\frac{e}{2M^{2}}\left(  \epsilon_{\gamma\delta\rho\alpha
}\partial^{\rho}T_{\beta}^{(\gamma\delta)}+\epsilon_{\gamma\delta\rho\beta
}\partial^{\rho}T_{\alpha}^{(\gamma\delta)}\right),  \label{uun}\\
&&T_{(\alpha\beta)\gamma}=-\frac{1}{2e}\epsilon_{\;\;\alpha\beta}
^{\rho\delta}\partial_{\rho}h_{\delta\gamma}.\label{ddo}
\end{eqnarray}
From Eq. (\ref{ddo}) and Eqs. (\ref{h1}-\ref{h2}) we obtain Eqs.
(\ref{k1}) and (\ref{k2}). Conversely, applying Eq. (\ref{uun})
to the expressions (\ref{k1}) and (\ref{k2}) we recover
(\ref{h1}) and (\ref{h2}). It is interesting to observe that the
term that diverges in the zero mass limit in (\ref{hmunu}) does
not contribute to $T_{(\alpha\beta)\gamma}$, which remains non
divergent in this limit. Thus the description in terms of
$T_{(\alpha\beta)\gamma}$ seems more suitable for studying the
massless limit.

The analysis of how the massless limit and the van
Dam-Veltman-Zakharov discontinuity appears in the dual theory
requires the discussion of duality in the case of $M=0$. We
postpone the discussion of this interesting point to a
forthcoming paper.

\section{Summary and final remarks}
In this article we have shown that a generalization of  duality
transformations applicable to arbitrary Lorentz tensors is
possible, and we have developed a scheme to construct such dual
descriptions. The main idea of this scheme is the use of a first
order parent Lagrangian from which either the original theory or
the dual one can be obtained, by means of permissible
substitutions arising from the algebraic solutions of the
corresponding equations of motion. This procedure guarantees the
equivalence of both theories and thus provides an adequate
dualization for arbitrary spin, either in the free or in the
coupled to an external source case, in contrast with previous
proposals \cite{t0}. In such a way one is able to construct new
non trivial and more general actions to describe  a given
physical system, which might show some advantages over the
standard ones.

Given a Lagrangian to be dualized, we first construct an
equivalent auxiliary first order Lagrangian written in  standard
form, which can always be done by using the method of Lagrange
multipliers in the manner described in Refs. \cite{Lanczos} and
\cite{MM}. This first order Lagrangian is not unique and provides
the identification of what we have called the field strength of
the original field. Dualization occurs at this level, through the
introduction of the dual tensor defined by the contraction of the
Levi-Civita tensor with the field strength. Different
possibilities might arise at this level which will produce
alternative dual theories.

Substitution of the  field strength in terms of the dual tensor in
the auxiliary first order Lagrangian produces the parent
Lagrangian, which is a functional of the original field
configuration together with the new dual field. On the one hand,
the elimination of the dual field from this Lagrangian, via its
equations of motion,  always takes us back to the original second
order theory. On the other hand, the elimination of the original
field from the parent Lagrangian defines the dual theory.

This dual tensor plays different roles in the massive and the
massless cases, because the duality transformation is singular in
the limit $ M\rightarrow 0$. The mapping between dual theories is
also very different according to these cases. For $M\neq0$ the
dual tensor turns out to be the basic configuration variables of
the dual theory and its definition in terms of the original field
strength provides the relation among the resultant theories. Here
the dual field is interpreted as a potential. For $M=0$ the
equation of motion of the dual field becomes a constraint (a
Bianchi identity) on the dual variable, which implies that it can
be written in terms of a potential. Hence the dual field can be
interpreted as a new field strength. The connection between both
theories is now  given  by a relation between the original and
dual potentials which usually involves derivatives. Summarizing,
for massive theories duality relates field strengths and
potentials, while for massless theories it relates the
corresponding potentials.

We have applied this scheme to the massive spin-2 field coupled to
external sources, obtaining a family of dual theories. The
starting point is the symmetric massive Fierz-Pauli field
$h_{\mu\nu}$ with its standard Lagrangian. The corresponding
first order auxiliary  Lagrangian, which  has two independent
parameters, is written in terms of $h_{\mu\nu}$  plus  the  field
strength $K_{\alpha\{\beta\gamma\}}$. The latter satisfies
additional symmetry properties. We have explicitly shown that the
elimination of the auxiliary field leads to the original  massive
Fierz-Pauli Lagrangian. At this stage there is some freedom in
the election of the dual field $\Omega$ and we have chosen the
relation $K^{\alpha\{\beta\gamma\}}=\epsilon^{\alpha\mu\nu\rho}\,
\Omega^{\{\beta\gamma\}}_{(\mu\nu\rho)}$. In order to partially
fulfill the induced symmetry properties of
$\Omega^{\{\beta\gamma\}}_{(\mu\nu\rho)}$ we have introduced the
auxiliary tensor $T^{(\alpha\beta)\gamma}$ which is required only
to be antisymmetric in the first two indices, and which serves as
the basic dual field in the sequel. This field is reminiscent of
what is called the Fierz tensor in Ref. \cite{NOVELLO}. Our
approach for the massive case is different from the latter
reference because we take $T^{(\alpha\beta)\gamma}$ as the basic
variable for the massive situation. The Lagrangian for this field
is subsequently constructed by eliminating $h^{\mu\nu}$ from the
parent Lagrangian. By construction this dual theory is equivalent
to the initial Fierz-Pauli description, and the connection between
both is established. The correct number of degrees of freedom in
the dual theory is verified by identifying the Lagrange
constraints arising from the equations of motion.

Finally, we have discussed  the case of the massive field
generated by  a point mass $m$ at rest, which is described using
both the original and the dual theory. This simple example
suggests that the description in terms of
$T^{(\alpha\beta)\gamma}$ behaves continuously in the limit
$M\rightarrow0$, in contrast with that in terms of $h^{\mu\nu}$.
The latter theory develops a singularity in the $M\rightarrow 0$
case , while the components of the dual field remain finite.

We postpone for a separate publication a detailed discussion of
the $M=0$  case. This situation is directly related to the
problem of dualizing linearized gravity, which has been the
subject of recent investigations
\cite{OBREGON},\cite{NIETO},\cite{HULL}. Our preliminary work on
this subject shows some interesting features: (i) the zero mass
limit of the dual Lagrangian for $T^{(\alpha\beta)\gamma}$, given
in Eq. (\ref{LAG}), has no arbitrary parameters so that one would
expect it to be completely determined by a set of gauge
symmetries to be determined. (ii) the Dirac analysis of the
constraints leads to the count of two degrees of freedom per
space point, in contrast with the results in Ref.\cite{t0}. The
analysis of the gauge structure that arises in the approach
pursued here should be of some interest, together with the
discussion of the Van Dam-Veltman-Zakharov discontinuity in the
dual theory.

\section*{Acknowledgments}
This work was partially supported by CONICET-Argentina and
CONACYT-M\'{e}xico. LFU acknowledges support from DGAPA-UNAM
project IN-117000, as well as CONACYT project 32431-E. He also
thanks the program CERN-CONACYT  for additional support.

\section*{Appendix A: First order Lagrangian in the standard form}

Here we show how to construct a first order Lagrangian equivalent
to a given second order Lagrangian using the method of Lagrange
multipliers. This approach has been presented in the framework of
classical mechanics for regular systems in Ref. \cite{Lanczos},
to construct a Hamiltonian formalism without the use of a
Legendre transformation. Ref. \cite{MM} deals with its application
to singular systems. To give a general idea of the method, we
simply sketch it for the case of regular field theories.
Consider  a given regular Lagrangian
\begin{equation}
L=L\left( \psi ^{a},\psi^{a}{}_{,\mu }\right).
\end{equation}
Let us assume that we want to introduce a set of new functions
$f_{\mu }^{a}=f_{\mu }^{a}(\psi^{b}{} _{,\mu })$, and treat them
as new independent fields. To do this it must be possible to solve
$\psi^{a}{}_{,\mu }$ in terms of $f_{\mu }^{a}$,.i.e. $\left|
\frac{\partial f_{\mu }^{a}}{\partial \psi _{,\nu }^{b}}\right|
\neq 0$. Under such a condition, we can write a new Lagrangian of
the form
\begin{equation}
L=L\left( \psi ^{a},f_{\mu }^{a}\right),
\end{equation}
and impose the constraints $f_{\mu }^{a}-f_{\mu }^{a}(\psi _{,\mu
}^{b})=0$. This can be done in a consistent way by introducing a set of
Lagrange multipliers $\lambda_a ^{\mu }$
\begin{equation}
\tilde{L}=L\left( \psi ^{a},f_{\mu }^{a}\right) +\lambda _{a}^{\mu }\left(
f_{\mu }^{a}(\psi _{,\mu }^{b})-f_{\mu }^{a}\right).
\end{equation}
The new formulation is clearly equivalent to the original one. The
auxiliary functions $f_{\mu }^{a}$ exclusively appear as
algebraic variables, without derivatives. Therefore, they can be
eliminated by using their equations of motion
\begin{equation}
\frac{\partial \tilde{L}}{\partial f_{\mu }^{a}}=\frac{\partial L}{\partial
f_{\mu }^{a}}-\lambda _{a}^{\mu }=0.
\end{equation}
The solution of this equation is a set of functions $f_{\mu
}^{a}=f_{\mu }^{a}(\psi ^{a},\lambda _{a}^{\mu })$,  and the
resulting Lagrangian has the form
\begin{equation}
\tilde{L}=\lambda _{a}^{\mu }f_{\mu }^{a}(\psi _{,\mu }^{b})-\lambda
_{a}^{\mu }f_{\mu }^{a}(\psi ^{a},\lambda _{a}^{\mu })+L\left( \psi
^{a},f_{\mu }^{a}(\psi ^{a},\lambda _{a}^{\mu })\right).
\end{equation}
The first term of the above Lagrangian shows that $\lambda
_{a}^{\mu }$ define the field strength of $\psi^b$. Their
relationship with the configuration variables is given by their
equation of motion in the first order theory
\begin{equation}
f_{\mu }^{a}(\psi _{,\mu }^{b})-f_{\mu }^{a}(\psi ^{a},\lambda
_{a}^{\mu })-\lambda _{b}^{\nu }\frac{\partial f_{\nu
}^{b}}{\partial \lambda _{a}^{\mu }}+\frac{\partial L}{\partial
\lambda _{a}^{\mu }}=0.
\end{equation}
This definition of the  field strength is not unique, because it depends on
the choice of the functions $f_{\mu }^{a}(\psi _{,\mu }^{b})$.

\section*{Appendix B: Equivalence between second order and first order Lagrangians}

In this Appendix we  show that Lagrangian (\ref{FOLFP}), where
the coupling constants satisfy Eqs. (\ref{cr1}) and (\ref{cr2}),
is a first order Lagrangian for the massive Fierz-Pauli theory
(\ref{FLP}). From Lagrangian (\ref{FOLFP}) we obtain the
equations of motion for $K_{\alpha \{\beta \sigma \}}$ and $
\Lambda _{\alpha \left\{ \beta \sigma \right\} }$
\begin{eqnarray}
&&-\frac{1}{3}aK_{\alpha \{\beta \sigma \}}+\frac{1}{4}q\,g_{\beta \sigma
}K_{\alpha }-\frac{2}{9}r\left( \epsilon _{\;\;\;\alpha \beta }^{\gamma
\delta }K_{\gamma \{\delta \sigma \}}+\epsilon _{\;\;\;\alpha \sigma
}^{\gamma \delta }K_{\gamma \{\delta \beta \}}\right) -\frac{e}{\sqrt{2}}
\partial _{\alpha }h_{\beta \sigma }+\Phi _{\alpha \beta \sigma }\,=0\,,
\label{EQMD} \\
&&K^{\alpha \{\beta \sigma \}}+K^{\beta \left\{ \sigma \alpha \right\}
}+K^{\sigma \left\{ \alpha \beta \right\} }=0\,,  \label{EQMDFP}
\end{eqnarray}
where
\begin{equation}
\Phi _{\alpha \beta \sigma }=\Lambda _{\alpha \left\{ \beta
\sigma \right\} }+\Lambda _{\beta \left\{ \sigma \alpha \right\}
}+\Lambda _{\sigma \left\{ \alpha \beta \right\} }\,,
\end{equation}
is a completely symmetric tensor. Eq. (\ref{EQMDFP}) leads to a
constraint between the two possible contractions of the indices
of $ K^{\alpha \left\{ \beta \sigma \right\} }$
\begin{equation}
\;\;K_{\alpha }+2K^{\beta }{}_{\left\{ \beta \alpha \right\}
}=0\,. \label{ck}
\end{equation}

First, we  solve the field strength $K_{\alpha \{\beta \gamma \}}$
in terms of the potential field\ $h_{\sigma \beta }$. We take the
two possible traces in Eq. (\ref{EQMD}), thus obtaining
\begin{eqnarray}
&&\frac{1}{3}\left(3q-a\right) K_{\alpha }-\frac{e}{\sqrt{2}}\partial
_{\alpha }h_{\beta }{}^{\beta }+\Phi _{\alpha \beta }{}^{\beta } =0\,, \\
&&\frac{1}{12}\left(3q+2a\right) K_{\alpha }-\frac{e}{\sqrt{2}}
\partial ^{\beta }h_{\beta \alpha }+\Phi _{\alpha \beta }{}^{\beta }
=0\,,
\end{eqnarray}
where we have used Eq. (\ref{ck}). Therefore
\begin{equation}
K_{\alpha }=\frac{2\sqrt{2}e}{\left(2a-3q\right) }\left( \partial
^{\beta }h_{\beta \alpha }-\partial _{\alpha }h_{\beta }{}^{\beta
}\right) \,.
\end{equation}
Performing a cyclic permutation of the indices in Eq.
(\ref{EQMD}) and adding the results we get
\begin{equation}
\Phi _{\alpha \beta \sigma }=\frac{e}{3\sqrt{2}}\left( \partial _{\alpha
}h_{\beta \sigma }+\partial _{\beta }h_{\sigma \alpha }+\partial _{\sigma
}h_{\alpha \beta }\right) -\frac{q}{12}\left( g_{\sigma \beta }K_{\alpha
}+g_{\beta \alpha }K_{\sigma }+g_{\alpha \sigma }K_{\beta }\right) \,.
\end{equation}
The contraction of Eq. (\ref
{EQMD}) with $\epsilon _{\;\;\;\mu \nu }^{\alpha \beta }$ gives
\begin{eqnarray}
&-&\frac{1}{3}a\epsilon _{\;\;\;\mu \nu }^{\gamma \delta
}\,K_{\gamma \{\delta \sigma \}}+\frac{1}{4}q\,\epsilon
_{\;\sigma \mu \nu }^{\alpha }K_{\alpha } +\frac{2}{3}r\left(
K_{\mu\{\nu \sigma \}}
-K_{\nu \{\mu \sigma \}}\right) \nonumber \\
&+&\frac{1}{3}r\left( g_{\mu \sigma }K_{\nu }-g_{\nu \sigma }K_{\mu }\right) -
\frac{e}{\sqrt{2}}\epsilon _{\;\;\;\mu \nu }^{\gamma \delta }\partial
_{\gamma }h_{\delta \sigma }=0\,.
\end{eqnarray}
Combining this last equation with Eq. (\ref{EQMD}) to eliminate
terms proportional to $\epsilon _{\;\;\;\mu \nu }^{\gamma \delta
}\,K_{\gamma \{\delta \sigma \}}$ we obtain the following
expression for $K_{\alpha \{\beta \sigma \}} $ in terms of
$h_{\alpha \beta }$
\begin{equation}
K_{\alpha \{\beta \sigma \}}=-\frac{\sqrt{2}}{e}\left[ A\,P_{\alpha \{\beta
\sigma \}}+B\,Q_{\alpha \{\beta \sigma \}}+C\,R_{\alpha \{\beta \sigma \}}
\right] \,,  \label{kaa}
\end{equation}
where
\begin{eqnarray}
\,P_{\alpha \{\beta \sigma \}} &=&g_{\alpha \sigma }\left(
\partial ^{\gamma }h_{\gamma \beta }-\partial _{\beta }h_{\gamma
}{}^{\gamma }\right) +g_{\alpha \beta }\left( \partial ^{\gamma
}h_{\gamma \sigma }-\partial _{\sigma }h_{\gamma }{}^{\gamma
}\right) -2g_{\beta \sigma }\left( \partial ^{\gamma
}h_{\gamma \alpha }-\partial _{\alpha }h_{\gamma }{}^{\gamma }\right) \,, \\
Q_{\alpha \{\beta \sigma \}} &=&\partial _{\beta }h_{\sigma \alpha
}+\partial _{\sigma }h_{\alpha \beta }-2\,\partial _{\alpha }h_{\beta \sigma
}\,, \\
R_{\alpha \{\beta \sigma \}} &=&\epsilon _{\;\;\;\alpha \beta }^{\gamma
\delta }\partial _{\gamma }h_{\delta \sigma }+\epsilon _{\;\;\;\alpha \sigma
}^{\gamma \delta }\partial _{\gamma }h_{\delta \beta }\,,
\end{eqnarray}
together  with the coefficients
\begin{eqnarray}
A &=&e^{2}\left( \frac{2}{3}r^{2}+\frac{1}{4}aq\right) \left( a-\frac{3}{2}
q\right) ^{-1}\left( a^{2}+4r^{2}\right) ^{-1}\,, \\
B &=&-\frac{1}{2}ae^{2}\left( a^{2}+4r^{2}\right) ^{-1}, \\
C &=&-re^{2}\left( a^{2}+4r^{2}\right) ^{-1}\,.
\end{eqnarray}
All three tensors appearing on the right hand side of Eq.
(\ref{kaa}) have a vanishing cyclic sum. The Lagrangian in terms
of $h_{\alpha \beta }$ can be obtained by replacing expression
(\ref{kaa}) in the first order Lagrangian (\ref{FOLFP}), or more
simply, noting that the contribution from the mass terms for
$K^{\alpha \{\beta \sigma \}}$ is $\left( -1/2\right)$ of the
contribution from the interaction term
$-\frac{e}{\sqrt{2}}K^{\alpha \{\beta \sigma \}}\partial _{\alpha
}h_{\beta \sigma }$. Half of this last contribution gives the
kinetic terms of the $h_{\alpha \beta }$ Lagrangian. The
interaction term gives
\begin{equation}
\left( 2A+2B\right) \partial _{\mu }h^{\mu \nu }\partial _{\alpha
}h_{\nu }{}^{\alpha }+\left( -2B\right) \partial _{\alpha }h^{\mu
\nu }\partial ^{\alpha }h_{\mu \nu }+\left( -4A\right) \partial
_{\mu }h^{\mu \nu }\partial _{\nu }h_{\alpha }{}^{\alpha }+\left(
2A\right) \partial _{\alpha }h_{\mu }{}^{\mu }\partial ^{\alpha
}h_{\nu }{}^{\nu }\,,
\end{equation}
after substituting Eq. (\ref{kaa}). Let us observe that the term
proportional to $C$ gives no contribution. Comparing  this
expression with the kinetic part of the original Lagrangian
(\ref{FLP}) we obtain $A=$ $(-1/2)$, $B=(-1/2)$. From here the
relations (\ref{cr1}) and (\ref{cr2}) follow.

\section*{Appendix C: Lagrangian constraints for the propagating field
$T^{(\protect\mu\protect\nu)\protect\rho}$}

We start from Eqs.(\ref{EQM}) and (\ref{EQM1}). The zero trace condition implies
\begin{equation}
F^{(\alpha \beta \theta )}{}_{\theta }=\partial _{\theta }T^{(\alpha \beta
)\theta }.
\end{equation}
We will also use the property
\begin{equation}
\epsilon _{\alpha \beta \gamma \delta }F^{(\beta \gamma \delta )\psi
}=3\epsilon _{\alpha \beta \gamma \delta }\partial ^{\beta }T^{(\gamma
\delta )\psi },
\end{equation}
together with the notation
\begin{equation}
T^{(\mu \nu \rho )}=T^{(\mu \nu )\rho }+T^{(\nu \rho )\mu
}+T^{(\rho \mu )\nu },
\end{equation}
and
\begin{equation}
D=\sqrt{a\left( e^{2}-a\right) }.
\end{equation}

Useful relations to determine the Lagrangian constraints are
obtained according to the following manipulations.

\begin{itemize}
\item  $g_{\gamma \nu }E^{\left( \beta \gamma \right) \nu }=0$ implies
\begin{equation}
\,\lambda ^{\beta }=\,\frac{4DM^{2}}{3e^{2}}\epsilon ^{\beta
\sigma \kappa \tau }T_{(\kappa \tau )\sigma }.  \label{LM}
\end{equation}

\item  {$\partial _{\beta }E^{(\beta \gamma )\nu }=0$ leads
to}
\begin{eqnarray}
&& {DM^{2}}\,\epsilon ^{\beta \gamma \kappa \lambda }{\partial
_{\beta }}T_{(\kappa \lambda )}{}^{\nu }-\frac{e^2}{4}\left(
g^{\gamma \nu }{
\partial _{\beta }}\lambda ^{\beta }-{\partial }^{\nu }\lambda ^{\gamma
}\right)   \nonumber \\
&& +\frac{2}{3}M^{2}\,{\partial _{\beta }}\left[ \left(
2a-\frac{1}{2} e^{2}\right) T^{(\beta \gamma )\nu
}+\frac{1}{2}\left( 2a+e^{2}\right) \left( T^{(\beta \nu )\gamma
}-T^{(\gamma \nu )\beta }\right) \right] =0. \label{DEQ1}
\end{eqnarray}
This expression can be decomposed in
the symmetric and antisymmetric part
\begin{eqnarray}
&&M^{2}{\partial _{\beta }}\left( T^{(\beta \gamma )\nu
}+T^{(\beta \nu )\gamma }\right) =-\frac{DM^{2}}{2a}\left(
\epsilon ^{\beta \gamma \kappa \lambda }{\partial _{\beta
}}T_{(\kappa \lambda )}{}^{\nu }+\epsilon ^{\beta \nu \kappa
\lambda }{\partial _{\beta }}T_{(\kappa \lambda )}{}^{\gamma
}\right) ,  \nonumber \\
&&\;\;\;\;\;\;\;\;\;\;\;\;\;\;\;\;\;\;\;\;\;\;\;\;\;\;\;\;\;\;\;\;\;\;+\frac{%
e^{2}}{4a}\left( g^{\gamma \nu }{\partial _{\beta }}\lambda ^{\beta }-\frac{1%
}{2}\left( {\partial }^{\nu }\lambda ^{\gamma }+{\partial }^{\gamma }\lambda
^{\nu }\right) \right)   \label{DEQ1S} \\
&&M^{2}\left( a-e^{2}\right) {\partial _{\beta }}\left( T^{(\beta \gamma
)\nu }-T^{(\beta \nu )\gamma }\right) =-\frac{3DM^{2}}{2}\left( \epsilon
^{\beta \gamma \kappa \lambda }{\partial _{\beta }}T_{(\kappa \lambda
)}{}^{\nu }-\epsilon ^{\beta \nu \kappa \lambda }{\partial _{\beta }}%
T_{(\kappa \lambda )}{}^{\gamma }\right)   \nonumber \\
&&\;\;\;\;\;\;\;\;\;\;\;\;\;\;\;\;\;\;\;\;\;\;\;\;\;\;\;\;\;+M^{2}{\partial
_{\beta }}\left( 2a+e^{2}\right) T^{(\gamma \nu )\beta }+\frac{3e^{2}}{8}%
\left( {\partial }^{\gamma }\lambda ^{\nu }-{\partial }^{\nu
}\lambda ^{\gamma }\right) .  \label{DEQ1A}
\end{eqnarray}

Applying $\partial _{\nu }$ to Eq. (\ref{DEQ1}) we  have
\begin{equation}
{4M^{2}}\,\left[ D\epsilon ^{\beta \gamma \kappa \lambda }\partial
_{\nu }{\partial _{\beta }}T_{(\kappa \lambda )}{}^{\nu
}+2a\partial _{\nu }{
\partial _{\beta }}T^{(\beta \gamma )\nu }\right] =e^2\,\partial _{\nu }\left( {
\partial }^{\gamma }\lambda ^{\nu }-{\partial }^{\nu }\lambda ^{\gamma
}\right).   \label{DADEQ1}
\end{equation}

\item  From $\epsilon _{\beta \gamma \nu \psi }E^{\left( \beta \gamma
\right) \nu }=0$ we obtain
\begin{equation}
2\,\epsilon _{\beta \gamma \nu \psi }\partial _{\alpha }F^{(\beta
\gamma \nu )\alpha }-6\,M^{2}\epsilon _{\beta \gamma \nu \psi }T^{(\beta \gamma
)\nu }-3\,\epsilon _{\beta \gamma \nu \psi }J^{(\beta \gamma )\nu
}=0.  \label{ee}
\end{equation}
Contracting  the above  equation
with $\epsilon ^{\rho \sigma \tau \psi }$ leads to
\begin{equation}
\partial _{\alpha }F^{(\beta \gamma \nu )\alpha }-M^{2}T^{(\beta \gamma \nu
)}=\frac{1}{2}\,J^{(\beta \gamma \nu )}.  \label{ee1}
\end{equation}

\item $\partial _\gamma E^{(\beta \nu )\gamma}=0$
implies
\begin{eqnarray}
\left( \partial ^{\nu }\lambda ^{\mu }-\partial ^{\mu }\lambda
^{\nu }\right)  &=&\frac{4DM^{2}}{e^{2}}\epsilon ^{\mu \nu \kappa
\lambda }\partial _{\theta }T_{(\kappa \lambda )}{}^{\theta
}+\frac{8aM^{2}}{
e^{2}}\partial _{\theta }T^{(\mu \nu )\theta }  \nonumber \\
&&+\frac{8D^{2}M^{2}}{3ae^{2}}\,\partial _{\theta }T^{(\theta \mu
\nu )}. \label{DEQ2}
\end{eqnarray}
\end{itemize}

This equation contains $\partial _{\theta }T^{(\mu \nu )\theta }$
and its dual $\frac{1}{2}\epsilon _{\alpha \beta \mu \nu
}\partial _{\theta }T^{(\mu \nu )\theta }$. We now show that this
relation leads to a solution of the combination
\begin{equation}
\Lambda _{\theta \beta }=\left( \partial _{\theta }\lambda
_{\beta }-\partial _{\beta }\lambda _{\theta }\right).
\end{equation}
Taking into account
that the constraint (\ref{LM}) gives for  $^{\ast }\Lambda ^{\mu \nu }=\frac{
1}{2}\epsilon ^{\mu \nu \theta \beta }\left( \partial _{\theta }\lambda
_{\beta }-\partial _{\beta }\lambda _{\theta }\right) $ the expression
\begin{equation}
^{\ast }\Lambda _{\theta \beta }=\frac{8DM^{2}}{3e^{2}}\partial _{\theta
}T^{(\theta \mu \nu )},
\end{equation}
we can rewrite Eq.(\ref{DEQ2}) in terms of $\Lambda ^{\mu \nu }$ and its
dual as
\begin{equation}
\Lambda ^{\mu \nu }+\frac{D}{a}\left( ^{\ast }\Lambda ^{\mu \nu }\right) =-
\frac{4DM^{2}}{e^{2}}\epsilon ^{\mu \nu \kappa \lambda }\partial _{\theta
}T_{(\kappa \lambda )}^{\;\;\;\;\;\theta }-\frac{8aM^{2}}{e^{2}}\partial
_{\theta }T^{(\mu \nu )\theta }.
\end{equation}
The dual of the above equation together with the property
$\,\,^{\ast }(^{\ast }\Lambda )=-\Lambda $ produce  a second
independent equation
\begin{equation}
\left( ^{\ast }\Lambda ^{\mu \nu }\right) -\frac{D}{a}\Lambda ^{\mu \nu }=
\frac{8DM^{2}}{e^{2}}\partial _{\theta }T^{\left( \mu \nu \right) \theta }-
\frac{4aM^{2}}{e^{2}}\epsilon _{\;\;\;\alpha \beta }^{\mu \nu }\partial
_{\theta }T^{(\alpha \beta )\theta }.
\end{equation}
Solving the system we are left with
\begin{eqnarray}
\Lambda ^{\mu \nu } &=&-\frac{8M^{2}a}{e^{2}}\partial _{\theta }T^{(\mu \nu
)\theta } , \label{DLAMB} \\
^{\ast }\Lambda ^{\mu \nu } &=&-\frac{4M^{2}a}{e^{2}}\epsilon
_{\;\;\;\alpha \beta }^{\mu \nu }\partial _{\theta }T^{(\alpha
\beta )\theta }. \label{DLAMB1}
\end{eqnarray}
These expressions are consistent with the duality relationship.

Eq. (\ref{DLAMB}) directly gives
\begin{equation}
M^{2}\partial _{\theta }T^{(\mu \nu )\theta }=-\frac{e^{2}}{8a}\Lambda ^{\mu
\nu }.  \label{DLAMB2}
\end{equation}

Taking now the divergence of Eq. (\ref{DLAMB1}) we obtain
\begin{equation}
\partial _{\mu }\left( \partial ^{\mu }\lambda ^{\nu }-\partial ^{\nu
}\lambda ^{\mu }\right) =-\frac{8aM^{2}}{e^{2}}\partial _{\mu }\partial
_{\theta }T^{(\mu \nu )\theta }.  \label{DDLAMB}
\end{equation}
The comparison of  the above relation with Eq. (\ref{DADEQ1}) gives
\begin{equation}
4M^{2}\epsilon ^{\beta \nu \kappa \lambda }\partial _{\sigma
}{\partial _{\beta }}T_{(\kappa \lambda )}{}^{\sigma }=0.
\end{equation}
Contracting the free index of this last equation with a Levi-Civita tensor
we obtain
\begin{equation}
4M^{2}\partial ^{\theta }F_{(\alpha \beta \kappa )\theta }=0,  \label{B1}
\end{equation}
which together with Eq. (\ref{ee1}) implies
\begin{equation}
M^{2}T^{(\beta \gamma \nu )}=-\frac{1}{2}\,J^{(\beta \gamma \nu )}.
\label{DT3}
\end{equation}
Taking the divergence of this equation respect to one of the antisymmetric
indices we get
\begin{equation}
M^{2}\partial _{\beta }\left( T^{(\beta \gamma )\nu }-T^{(\beta \nu )\gamma
}\right) =-M^{2}\partial _{\beta }T^{(\gamma \nu )\beta }-\frac{1}{2}
\,\partial _{\beta }J^{(\gamma \nu )\beta }.  \label{DT4}
\end{equation}
Using Eq. (\ref{DT3}) in Eq. (\ref{LM}) we obtain a relationship between the
Lagrange multiplier $\lambda _{\sigma }$ and the source:
\begin{equation}
\,\lambda ^{\beta }=-\,\frac{2D}{9e^{2}}\epsilon ^{\beta \kappa
\tau \sigma }J_{(\kappa \tau \sigma )}.
\end{equation}

Finally, Eqs. (\ref{DT4}) and (\ref{DLAMB2}) imply
\begin{equation}
M^{2}\partial _{\beta }\left( T^{(\beta \gamma )\nu }-T^{(\beta
\nu )\gamma }\right) =\frac{e^{2}}{8a}\Lambda ^{\gamma \nu
}-\frac{1}{2}\partial _{\alpha }J^{(\gamma \nu )\alpha }.
\label{SIMDT33}
\end{equation}

From the above results the following independent Lagrangian
constraints arise:
\begin{eqnarray}
&&\,\lambda _{\sigma }=-\frac{2D}{9e^{2}}\epsilon _{\sigma \beta \gamma \nu
}\,J^{(\beta \gamma \nu )},  \label{CC1} \\
&&M^{2}T^{(\beta \gamma \nu )}=-\frac{1}{2}\,J^{(\beta \gamma \nu
)},
\label{CC2} \\
&&M^{2}\partial _{\theta }T^{(\mu \nu )\theta }=-\frac{e^{2}}{8a}\Lambda
^{\mu \nu },  \label{CC3} \\
&&M^{2}{\partial _{\beta }}T^{(\beta \gamma )\nu }=\frac{1}{16a}\left[
e^{2}\Lambda ^{\gamma \nu }-4a\partial _{\alpha }J^{(\gamma \nu )\alpha }
\right]   \nonumber \\
&&+\frac{e^{2}}{8a}\left[ g^{\gamma \nu
}{\partial_\rho}\lambda^\rho -\frac{1}{2} \left( {\partial }^{\nu
}\lambda ^{\gamma }+{\partial }^{\gamma }\lambda ^{\nu }\right)
-2\frac{DM^{2}}{e^{2}}{\partial _{\beta }}\left( \epsilon ^{\beta
\gamma \kappa \lambda }T_{(\kappa \lambda )}{}^{\nu }+\epsilon
^{\beta \nu \kappa \lambda }T_{(\kappa \lambda )}{}^{\gamma
}\right) \right], \label{CC4}
\end{eqnarray}
and the equation of motion is
\begin{eqnarray}
&&2\partial _{\alpha }\partial ^{\alpha }T^{(\beta \gamma )\nu
}+\frac{D}{2a} {\partial _{\sigma }}\left[ \partial ^{\beta
}\left( \epsilon ^{\sigma \gamma \kappa \lambda }T_{(\kappa
\lambda )}{}^{\nu }+\epsilon ^{\sigma \nu \kappa \lambda
}T_{(\kappa \lambda )}{}^{\gamma }\right) -\partial ^{\gamma
}(\epsilon ^{\sigma \beta \kappa \lambda }T_{(\kappa \lambda
)}{}^{\nu }+\epsilon ^{\sigma \nu \kappa \lambda }T_{(\kappa
\lambda )}{}^{\beta })
\right] \nonumber \\
&&+\frac{M^{2}}{e^{2}}\sqrt{a\left( e^{2}-a\right) }\epsilon
^{\beta \gamma \kappa \lambda }T_{(\kappa \lambda )}{}^{\nu
}+\frac{2aM^{2}}{e^{2}}T^{(\beta \gamma )\nu
}=\frac{1}{2}\,J^{(\beta \gamma )\nu }+\frac{2a+e^{2}}{6e^{2}}
J^{(\gamma \beta \nu )} \nonumber\\
&&+\frac{1}{6M^{2}}\left( 2\partial _{\alpha }\partial ^{\alpha }J^{\left(
\gamma \beta \nu \right) }+2\partial ^{\nu }\partial _{\alpha }J^{\left(
\beta \gamma \right) \alpha }+\partial ^{\beta }\partial _{\alpha }J^{(\nu
\gamma )\alpha }+\partial ^{\gamma }\partial _{\alpha }J^{\left( \beta \nu
\right) \alpha }\right) \nonumber\\
&&+ \frac{1}{4}\left( g^{\gamma \nu }\lambda ^{\beta }-g^{\beta \nu }\lambda
^{\gamma }\right) +\frac{e^{2}}{4M^{2}a}\partial ^{\beta }\left( g^{\gamma
\nu }{\partial_\rho}\lambda^\rho -\frac{1}{2}{\partial }^{\nu }\lambda ^{\gamma
}\right) -\frac{e^{2}}{4M^{2}a}\left( g^{\beta \nu }\partial ^{\gamma
}-g^{\gamma \nu }\partial ^{\beta }\right) {\partial_\rho}\lambda^\rho \nonumber\\
&&-\frac{e^{2}}{24aM^{2}}\left( 8\partial ^{\nu }\Lambda ^{\beta
\gamma }+\partial ^{\beta }\Lambda ^{\nu \gamma }+\partial
^{\gamma }\Lambda ^{\beta \nu }\right)
-\frac{e^{2}}{8aM^{2}}\partial _{\alpha }\left( g^{\gamma \nu
}\Lambda ^{\alpha \beta }+g^{\beta \nu }\Lambda ^{\gamma \alpha
}\right).
\end{eqnarray}

In the case  $a=e^{2}$, we have
\begin{eqnarray}
&&\lambda ^{\beta }=\,0, \qquad \qquad \qquad
M^{2}T^{(\beta \gamma \nu )}=-\frac{1}{2}\,J^{(\beta \gamma \nu )}, \\
&&M^{2}\partial _{\theta }T^{(\mu \nu )\theta }=0, \qquad
M^{2}{\partial _{\beta }}T^{(\beta \gamma )\nu
}=-\frac{1}{4}\partial _{\alpha }J^{(\gamma \nu )\alpha },
\end{eqnarray}
\begin{eqnarray}
\label{EQTE2A}
\left(\partial _{\alpha }\partial ^{\alpha }+M^{2}\right)T^{(\beta
\gamma )\nu }&=&\frac{1}{4}\,J^{(\beta \gamma )\nu }+\frac{1}{4}
J^{(\gamma \beta \nu )} \nonumber \\
&&+\frac{1}{12M^{2}}\left( 2\partial _{\alpha }\partial ^{\alpha
}J^{\left( \gamma \beta \nu \right) }+2\partial ^{\nu }\partial
_{\alpha }J^{\left( \beta \gamma \right) \alpha }+\partial
^{\beta }\partial _{\alpha }J^{(\nu \gamma )\alpha }+\partial
^{\gamma }\partial _{\alpha }J^{\left( \beta \nu \right) \alpha
}\right).
\end{eqnarray}

\end{document}